\documentclass{ccjnl}

\title{Improved ALOHA-based URA with Index Modulation: Efficient Decoding and Analysis}
\author{Linjie Yang\inst{1,3}, Pingzhi Fan\inst{1,*}, Zhiguo Ding\inst{2}, and Jingqiu Gao\inst{1}
\corinfo{pzfan@swjtu.edu.cn} }
\receiveddate{}
\reviseddate{}
\Editor{}
\address[1]{School of Information Science and Technology, Southwest Jiaotong University, Chengdu 610031, China.}
\address[2]{Department of Electrical Engineering and Computer Science, Khalifa University, Abu Dhabi, UAE.}
\address[3]{Intelligent TT\&C and space-based application laboratory, The tenth research Institute of China electronics technology group corporation, China.}
\address[~]{This paper is an extended version of the work originally presented at the 2023 IEEE 15th International Conference on Wireless Communications and Signal Processing (WCSP). It has been accepted for publication in China Communications.}

\begin{document}
	\maketitle

\begin{abstract}
In this paper, an improved ALOHA-based unsourced random access (URA) scheme is proposed in MIMO channels.
The channel coherent interval is divided into multiple sub-slots and each active user selects several sub-slots to send its codeword, namely, the channel access pattern.
To be more specific, the data stream of each active user is divided into three parts.
The first part is mapped as the compressed sensing (CS) pilot, which also serves for the consequent channel estimation.
The second part is modulated by binary phase shift keying (BPSK).
The obtained CS pilot and the antipodal BPSK signal are concatenated as its codeword.
After that, the codeword of each active user is sent repeatedly based on its channel access pattern, which is determined by the third part of the information bits, namely, index modulation (IM).
On the receiver side, a hard decision-based decoder is proposed which includes the CS decoder, maximal likelihood (ML)-based superposed codeword decomposer (SCD), and IM demodulator.
To further reduce the complexity of the proposed decoder, a simplified SCD based on convex approximation is considered.
The performance analysis is also provided.
The exhaustive computer simulations confirm the superiority of our proposal.
\end{abstract}

\section{introduction}
\label{s1}

{Massive} machine type communication (mMTC)  has attracted comprehensive research interest in recent years, which has three main specific features: (1) the number of potential users is large (e.g., $10^{6}$ devices per ${\rm km}^{2}$); (2) the cellular users implement the sporadic transmission (i.e, the active user number is small in any transmission round); (3) the transmitted information bit length of active users is short (e.g., $\leq 100$ bits) \cite{bockelmann2018towards}.
The massive connectivity property of mMTC scenarios brings a challenge to the traditional orthogonal multiple access (OMA) structure, since the limitation of orthogonal user-specific pilot construction and the high overhead caused by the handshake procedure in the traditional random access (RA) process.
To address this dilemma, some researchers advocated assigning the non-orthogonal pilot to the users in the cell.
The active user detection (AUD) and channel estimation (CE) can be jointly formulated as a standard compressed sensing (CS) problem, which is efficiently solved by the approximate message passing algorithm (AMP) \cite{chen2018sparse,liu2018massive,liu2018sparse}.
Meanwhile, since the active users access the base station (BS) in a grant-free (GF) manner, the overhead in RA is significantly reduced.
However, the above-mentioned CS-based GF schemes generally support only thousands of potential users with the typical system configurations.
When the potential user number is massive, assigning non-orthogonal pilot to each cellular user becomes impractical.
Thus, designing the RA scheme which supports the ultra-massive connectivity in the 6G communication system, becomes an open problem.

In this regard, Polyanskiy et al. proposed the concept of unsourced random access (URA), where each active user generates codeword from the same codebook and the decoder's task is to estimate the list of transmitted messages regardless of the permutation of messages \cite{Polyanskiy-perspective}.
In 2017, Ordentlich et al. proposed the first practical URA scheme, which is referred to as T-fold ALOHA under Gaussian channel \cite{original-T-fold-ALOHA}. 
This approach is similar to the standard slotted-ALOHA where channel uses are split into sub-blocks and each active user would randomly select a sub-block for transmission.
The main difference is that, in T-fold ALOHA, if the superposed codewords are no more than $T$ during the same sub-block, all corresponding messages can be decoded.
To this end, a concatenated coding scheme is proposed, where the bit stream of each active user would be firstly encoded by a binary outer code, subsequently, the obtained codeword is further encoded by a linear inner code.
Kowshik et al. analyzed the achievable bound of T-fold ALOHA under the Rayleigh channel in \cite{T-fold-ALOHA-rayleigh}.
Ustinova et al. proposed T-fold irregular repetition slotted ALOHA (IRSA) \cite{T-fold-IRSA}, where each active user selects several sub-blocks for transmission based on a predesigned degree distribution. Furthermore, to further improve the throughput, the successive interference cancellation (SIC) procedure is considered.
In \cite{analysis1-T-fold-IRSA}, Glebov et al. analyzed the achievable bound of T-fold IRSA by combining the density evolution method and a finite-length random coding bound proposed by Polyanskiy.
In \cite{analysis2-T-fold-IRSA}, Ustinova et al. derived the optimal repetition distribution with the condition of different $T$ and the fixed error probability. 
In \cite{CS+sparse-spreading}, the {authors} divided the bit stream into two parts, where the first part is mapped as a compressed sensing (CS) pilot indicating a specific interleaving sequence. The second part is encoded by the low-density-parity-check (LDPC) code whose permutation is determined by the aforementioned interleaving sequence.
In \cite{polar+T-fold},
{Andreev et al.} replaced the LDPC code-based inner code of T-fold IRSA \cite{CS+sparse-spreading} with Polar code and showed that TIN-SIC decoding significantly outperforms the joint decoder given in \cite{CS+sparse-spreading}.
Since these T-fold ALOHA-based URA schemes can be expressed efficiently by the Tanner graph, we referred to this type of scheme as sparse graph-based URA schemes.

Indeed, URA has a strong connection with the sparse vector recovery problem in compressed sensing (CS) scenarios.
Unfortunately, the dimension of the common codebook would increase prohibitively when the transmitted bit stream increases to an order of hundred bits.
To tackle this issue, 
Amalladinne et al.
introduced the coded compressed sensing (CCS) framework, where the divide-and-conquer principle is employed \cite{amalladinne2020-original-CCS}.
In CCS,  
the message of each active user is divided into several sub-blocks such that the common codebook size can be reduced greatly.
Besides, an outer tree code is employed to build the connections between these sub-blocks, based on which the detected data pieces can be stitched together.
In \cite{fengler2021sparcs}, Fengler et al. adopted sparse regression code (SPARC) to act as the inner code of CCS under the Gaussian channel. On the receiver side, a modified approximate message passing (AMP) algorithm is developed to be the inner decoder, whose error probability is analyzed precisely.
Based on \cite{fengler2021sparcs}, Amalladinne et al. pointed out that the redundancy intrinsic to the outer code can be utilized to accelerate the convergence of the AMP algorithm.
In this spirit, a modified outer code enjoying the low complexity is developed \cite{IntegrateAMP}.
Besides, the outer code construction of CCS is a concern for some researchers \cite{cao2023crc,CCS_list_code,jiang2023raptor}.
In \cite{cao2023crc}, the SPARC and CRC are considered in the encoding process. 
In \cite{CCS_list_code}, Andreev et al. constructed two types of outer code of CCS which can correct $t$ errors. 
In another line of CCS studies, some researchers attempt to avoid the utilization of the explicit outer code in CCS to improve the spectral efficiency. 
Consequently, the decoded data pieces from different sub-blocks are stitched through some implicit hints \cite{uncoupledCCS,CCSangulardomain,CCSbeamspace,CCSRM}.
In \cite{uncoupledCCS}, the user-specific channel coefficients is exploited. 
The unique transmission feature in angular domain is utilized in \cite{CCSangulardomain}.
In \cite{CCSbeamspace}, Jingze et al. proposed a novel URA scheme based on beam-space tree decoding for millimeter wave (mmWave) massive MIMO system. 
In \cite{CCSRM}, Jue et al. adopted the novel shift property of second order Reed-Muller (RM) sequences to realize the stitching process of detected data pieces.

Besides, there are also some hybrid URA schemes \cite{fengler2022pilot,multiple-stage-orthogonal-pilot,FASURA}.
In \cite{fengler2022pilot}, Fengler et al. proposed a pilot-based URA scheme where the message of the active user is divided into two parts.
The first part is encoded by the CS encoder, which also undertakes the channel estimation (CE).
The second part is encoded by Polar code.
At the receiver, based on the estimated channel coefficients of the CS solver, the superposed codeword of the second part of the transmitted messages is decomposed by the maximal ratio combination (MRC) algorithm.
In \cite{multiple-stage-orthogonal-pilot}, multiple stages orthogonal pilot is introduced into Fengler's scheme \cite{fengler2022pilot}. 
In \cite{FASURA}, Gkagkos et al. first employed multiple spreading modes to further improve the decoding performance at the expense of the higher receiver complexity. 
In \cite{slotted_pilot}, Ozates et al. improved the performance of the pilot-based URA scheme \cite{fengler2022pilot} by introducing multiple slot to reduce the multi-user interference.

It is noted, among the most of sparse graph-based URA schemes, the channel estimation is generally ignored.
On the other hand, the receivers of the CCS-based URA and the hybrid URA are complex, due to the advanced channel coding utilization. 
{Thirdly, the existed URA schemes, e.g. the sparse-graph based URA schemes, CCS-based URA scheme and the hybrid URA schemes, are mainly designed for the quasi-static Internet-of-things (IoT) devices and always require a long coherent channel duration.
It implies that these schemes are inapplicable to support
the devices with a certain velocity (e.g. the agricultural robot, the automated guided vehicle, and the intelligent wearable device). }
To this this context, Jiaai et al. proposed a novel sparse graph-based URA approach \cite{liu2022unsourced}, where the channel estimation is realized with the extremely low channel consumption.
To be more concrete, the transmitted information bits of an active user are purely modulated by the binary phase shift keying (BPSK) modulation.
Then, a single `+1' symbol is added to be the header of the codeword for channel estimation purpose.
Apparently, such an approach reduces the codeword length significantly, which results in a higher spectral efficiency as well as an easier hardware realization.
To our best knowledge, the method in \cite{liu2022unsourced} has the shortest codeword length, which implies that it is also more 
suitable to the fast fading scenarios \cite{multiple-stage-orthogonal-pilot} (The coherence block length $L_{c} \leq 300$\cite{multiple-stage-orthogonal-pilot}).
However, limited by the proposed single symbol-based channel estimation method, the complexity of the decoder rapidly becomes prohibitive upon the collided codeword number on the same sub-slot increasing (e.g., only up to 3 collided codewords at the same sub-slot are supported).
Hence, the spare graph-based URA method \cite{liu2022unsourced} would be inefficient in the dense active user scenarios.

In this paper, an improved ALOHA-based URA scheme is proposed based on the insights in \cite{liu2022unsourced}. 
{
In this sense, this work is an improvement of the sparse-graph-based method in \cite{liu2022unsourced}.
The main differences lies in two aspects:
	\begin{itemize}
		\item Inspired by the success of \cite{fengler2022pilot,uncoupledCCS}, the spatial diversity in the MIMO channel is exploited to distinguish the collided codewords on the same sub-slot. Similar to \cite{fengler2022pilot,polar+T-fold_MIMO}, we adopted the CS pilot for CE as well as conveying some information bits, based on which the consequent BPSK modulated signal can be decomposed. 
		\item To circumvent the channel use consumption introduced by the CS pilot, the indexes of sub-blocks where the active users send their codewords (i.e., channel access pattern) are further explored to convey information bits implicitly, {which is referred to as index modulation (IM). 
		IM schemes generally map the information bits by altering the on/off status of their transmission entities such as transmit antennas, sub-carriers, radio frequency (RF) mirrors, modulation types, time slots, spreading codes and so on \cite{mao2021terahertz,mao2024index,xiao2023twin}.
		In our proposal, each active user would pick $K$ out of $N_{slot}$ slots for transmission. The length of information bits conveyed by IM is $L_{bI} = \lfloor \log_{2}(\binom{N_{slot}}{K}) \rfloor$. Assume the length of information bits conveyed by the CS pilot is $L_{bp}$, the length of the CS pilot $L_p = L_{bI}+L_{bp}$. 
		By such an approach, the channel estimation performance can be significantly improved.
		Meanwhile,}
		the codeword length in our proposal keeps the same as that in \cite{liu2022unsourced}.
	\end{itemize}
}

Our main contributions can be briefly summarized as follows.
\begin{itemize}
	\item An improved ALOHA-based URA scheme is proposed, where the information bits are divided into three bit streams. The first bit stream is conveyed by the CS encoding, which also serves for the consequent CE. The second bit stream is  modulated by BPSK. The third information part determines the sub-slots which are employed for the data transmission, i.e., index modulation (IM).
	
	\item {A hard decision (HD)-based decoder is proposed to implement the data decoding, which includes the CS decoder, the maximum likelihood-based superposed codeword decomposer (ML-SCD) and the IM demodulator.
		To further reduce the complexity of the proposed decoder, a simplified SCD based on convex approximation is considered.
		In addition, the CS pilot collision resolution of the proposed scheme is also discussed.}
	
	\item {To guarantee the execution of the successive interference cancellation (SIC), we analyze the probability that there is at least a decodable sub-slot, based on which the proper sub-slot number can be selected.
		Moreover, the throughput of the proposed scheme is also analyzed through the density evolution tool developed in \cite{liu2022unsourced}.
		In addition, the complexity analysis of the proposed decoder is also provided.}
	
	\item {Finally, the simulation results reveal that, compared with the counterparts, the proposed scheme can support around $115$ active users at the target frame error rate (FER) equals $0.05$, whereas the sparse graph-based URA scheme \cite{liu2022unsourced} only supports around $60$ active users. Accordingly, the proposed scheme also achieves a higher throughput.}
	
\end{itemize}

The rest of the paper is organized as follows.
The system model is described in detail in Section II.
The HD-based decoder design is elaborated in Section III.
Two simplfied superposed codeword decomposers are given in Section IV.
The simulation results are presented in Section V.
The paper is concluded in Section VI.

\emph{\textbf{Notations}:}
Throughout this paper, scalars are represented in lowercase letters.
Boldface lowercase and uppercase letters denote vectors and matrices, respectively.
The length-N vectors of all zeros and all ones are denoted as $0_{N}$ and $1_{N}$, respectively.
The notation $(\cdot)^{T}$ means the transpose operation.
Similarly, the notation $(\cdot)^{H}$ denotes the conjugate transpose operation. $\vert \mathcal{A}\vert$ denotes the number of elements in the set $\mathcal{A}$.
$\mathbf{b}[m]$ denotes the $m$-th element of the vector $\mathbf{b}$.
$\mathbf{B}[m,n]$ denotes the element in the $m$-th row and the $n$-th column of the matrix $\mathbf{B}$.
$\mathbf{I}_{d}$ denotes the $d\times d$ identity matrix.
We denote a diagnal matrix with elements $(s_{1}, s_{2}, \cdots, s_{k})$ by $\rm diag(s_{1}, s_{2},\cdots,s_{k})$.

\section{System Model}
\label{s2}

An up-link scenario is considered where a single base station (BS) serves $N_{tot}$ potential users.
These potential users access the BS in a sporadic manner, i.e., the active user number $N_{a} \ll N_{tot}$ in any given transmission period.
In addition, the BS is equipped with $M$ antennas and the potential users are equipped with a single antenna.
When the $u$-th $1\leq u \leq N_{tot}$ user become active, its activity state would be changed from $a_{u} = 0$ to $a_{u} = 1$.
Then, each active user would modulate its bit stream  $\mathbf{b}_{u}$ as the transmitted signal $\mathbf{x}_{u}$ through a modulation process.
The length of $\mathbf{b}_{u}$ is denoted as $L_{bs}$, i.e.,
$\mathbf{b}_{u} \in \{0,1\}^{L_{bs}\times 1}$.

\begin{figure*}[htbp]
	\centering
	\includegraphics[scale=1]{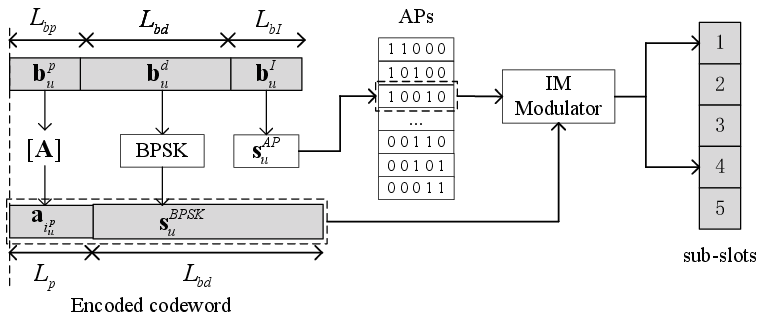}
	\caption{Illustration of the proposed hybrid encoding process with index modulation, where $N_{slot}=5, K = 2$.   }
	\label{system_model}
	\vskip -3mm
\end{figure*}

The encoding process of the proposed scheme is given in Fig. \ref{system_model}, which can be roughly divided into two steps, namely message split step and the IM step, respectively.
As is shown in Fig. \ref{system_model}, the bit stream of the active user $u$, $\mathbf{b}_{u}$ is divided into three parts, namely the CS pilot part, $\mathbf{b}_{u}^{p}$, the BPSK data part, $\mathbf{b}_{u}^{d}$ and the index modulation (IM) part, $\mathbf{b}_{u}^{I}$.
The lengths of $\mathbf{b}_{u}^{p}$, $\mathbf{b}_{u}^{d}$ and $\mathbf{b}_{u}^{I}$ are denoted as $L_{bp}$, $L_{bd}$ and $L_{bI}$ respectively.
Then, the first part of information bits $\mathbf{b}_{u}^{p}$ is mapped into the $i_{u}^{p}$-th CS pilot of a common CS pilot codebook $\mathbf{A}=[\mathbf{a}_{1},\mathbf{a}_{2},\cdots,\mathbf{a}_{2^{L_{bp}}}]$, $\mathbf{a}_{i_{u}^{p}}$ {whose length is $L_p$}. 
The CS codebook $\mathbf{A}$ is constructed by the partial Hadamard matrix method with a low mutual coherence value \cite{yang2023improved}.
The $i_{u}^{p}$ is calculated by
\begin{equation}
	i_{u}^{p} = {\rm dec}(\mathbf{b}_{u}^{p})+1,
\end{equation}
where the function {${\rm dec}(\mathbf{b}_{u}^{p})$} returns the decimal form of the binary vector $\mathbf{b}_{u}^{p}$.
The second part of information bits $\mathbf{b}_{u}^{d}$ is modulated by BPSK modulation, which is formulated as
\begin{equation}
	\mathbf{s}_{u}^{\rm BPSK} = 2\mathbf{b}_{u}^{d}-1.
\end{equation}
{The length of $\mathbf{s}_{u}^{\rm BPSK}$ is identical to $L_{bd}$.}
Afterwards, the codeword of active user $u$, $\mathbf{x}_{u}$ is constructed by concatenating the CS pilot and the antipodal BPSK signal, which is given by
\begin{equation}
	\mathbf{c}_{u} = [\mathbf{a}_{i_{u}^{p}}, \mathbf{s}_{u}^{\rm BPSK}],
\end{equation}
{The length of $\mathbf{c}_{u}$ is $L_{bs}+1$}.
After that, the third part of information bits is mapped as a channel access pattern sequence, $\mathbf{s}_{u}^{AP}$.
Note that $\mathbf{s}_{u}^{AP}$ is a binary vector where the number of element one is $K$ (e.g., $K=2$).
{The length of $\mathbf{s}_{u}^{AP}$ is identical to the number of sub-slot, $N_{slot}$.}
Note that there are $ \binom{ N_{slot} }{K}$ channel access pattern sequences (APs). 
{
We can arrange these channel access pattern sequences row-by-row and obtain the APS pool, $\mathbf{H}\in\{0,1\}^{\binom{N_{slot}}{K} \times N_{slot}  }$. 
$\mathbf{s}_{u}^{AP}$ is given by
\begin{equation}
	\mathbf{s}_{u}^{AP} = \mathbf{H}[{\rm dec}(\mathbf{b}_{u}^{I}),  :].
\end{equation}
For example, $N_{slot}=5, K=2$, $\mathbf{H}^{T}$ is given by
\begin{equation*}
	\mathbf{H}^{T} = 
	\begin{pmatrix}
		1 & 1 & 1 & 1 & 0 & 0 & 0 & 0 & 0 & 0 \\
		1 & 0 & 0 & 0 & 1 & 1 & 1 & 0 & 0 & 0 \\
		0 & 1 & 0 & 0 & 1 & 0 & 0 & 1 & 1 & 0 \\
		0 & 0 & 1 & 0 & 0 & 1 & 0 & 1 & 0 & 1 \\
		0 & 0 & 0 & 1 & 0 & 0 & 1 & 0 & 1 & 1 
	\end{pmatrix}.
\end{equation*}  
}
Furthermore, the information bits conveyed by IM approach, $L_{bI}$ can be computed as
\begin{equation}
	L_{bI} = \lfloor \log_{2}( \binom{N_{slot}}{K}  )   \rfloor.
\end{equation}
Such an IM approach is widely adopted in the sparse vector coding (SVC), which is a promising technique for the extremely short packet transmission in the ultra-reliable and low-latency communication (URLLC) scenarios \cite{SSC}. 
Similar to \cite{fengler2022pilot,slotted_pilot}, $L_{bp}$ is chosen such that the collision probability of the CS pilot is extremely low (e.g., $L_{bp}=14$ \cite{fengler2022pilot}) in this paper. 
Hence, the value of $L_{bd}$ is given by
\begin{equation}
	L_{bd} = L_{bs} - L_{bp} - L_{bI}.
\end{equation}
Finally, active user $u$ would send its codeword on different sub-slots based on its channel access pattern sequence $\mathbf{s}_{u}^{AP}$.
In Fig. \ref{system_model}, a toy example of the spreading process is given where $N_{slot}=5, K=2$.
The transmitted signal of active user $u$ is formulated as
\begin{equation}
	\mathbf{x}_{u} = \mathbf{s}_{u}^{AP}\otimes \mathbf{c}_{u},
\end{equation}
where $\otimes$ denotes the Kronecker product of two vectors.
The channel use consumption is $N_{cu}=N_{slot}(L_{bs}+1)$.
The transmitting rate is $r=\frac{N_{a}L_{bs}}{ N_{slot}(L_{p}+L_{\rm BPSK})  } = \frac{N_{a}L_{bs}}{ N_{slot}(L_{bs}+1)  } \approx \frac{N_{a}}{N_{slot}}  $.
Denote $\mathbf{g}_{u}\in \mathbb{C}^{M}$ as the channel vector between active user $u$ and the BS, the received signal at the BS can be formulated as
\begin{equation}
	\mathbf{Y} = [\mathbf{Y}_{1},\cdots,\mathbf{Y}_{N_{slot}}] = \sum_{u=1}^{N_{a}}\mathbf{g}_{u}\mathbf{x}_{u}^{T} + \mathbf{n} \in \mathbb{C}^{M\times N_{cu}}.
\end{equation}
$\mathbf{n}\in \mathbb{C}^{M\times N_{cu}}$ is the background Gaussian noise which obeys to $\mathcal{CN}(0,\sigma^{2})$.
The channel coefficients of users $\mathbf{g}[m], 1\leq m \leq M$ are i.i.d. and obeys to the one-dimension complex Gaussian distribution $\mathcal{CN}(0,1)$.
$\mathbf{Y}_{n_{s}}\in \mathbb{C}^{M\times (L_{bs}+1)}, 1\le n_{s}\leq N_{slot}$ is the received signal during the $n_{s}$-th sub-slot.
Similarly, $\mathbf{n}_{n_{s}}$ denotes the Gaussian noise on the $n_{s}$-th sub-slot.

\begin{figure*}[htbp]
	\centering
	\includegraphics[scale=0.74]{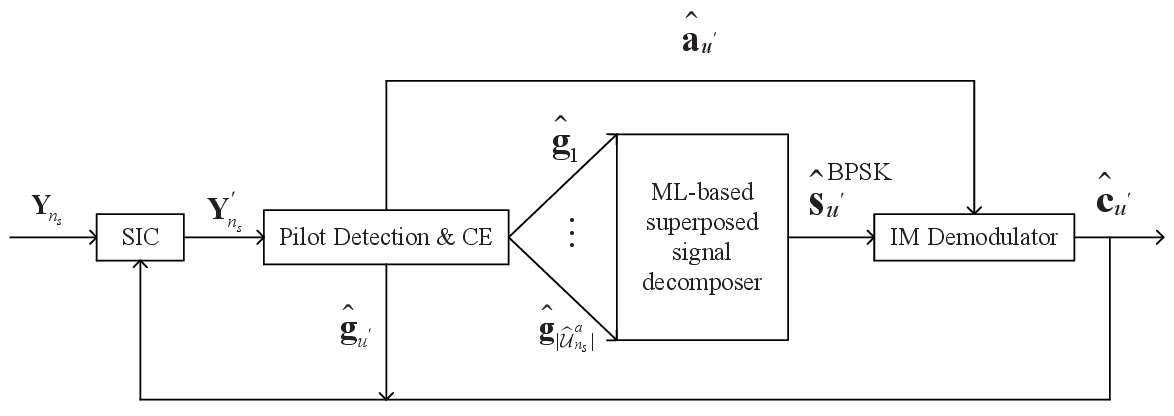}
	\caption{The decoding process of the proposed scheme on the $n_{s}$-th sub-slot.}
	\label{decoder}
	\vskip -3mm
\end{figure*}

\section{Hard-decision-based Decoder}
\label{s3}
At the receiver, the data decoding is implemented on the slot-wise.
In this section, we will elaborate on the hard-decision (HD)-based decoder design, which includes the CS decoder, the maximum-likelihood (ML)-based superposed codeword decomposer and the IM demodulator.

\subsection{CS pilot detection \& CE}
For convenience, we focus on the CS pilot detection of the $n_{s}$-th sub-slot.
The received signal of the CS pilot part on the $n_{s}$-th sub-slot, $\mathbf{Y}_{n_{s}}^{p} = \mathbf{Y}_{n_{s}}[:,1:L_{p}] $, at the BS is given by
\begin{equation}
	\mathbf{Y}_{n_{s}}^{p}= \sum_{i_{ u^{\prime}} \in \mathcal{U}_{n_{s}}^{a}} \mathbf{a}_{i_{u'}} \mathbf{g}_{u'}^{T} + \mathbf{n}_{n_{s}}^{p} \in \mathbb{C}^{M\times L_{p}},
\end{equation}
where $\mathcal{U}_{n_{s}}^{a}$ is the index set of the CS pilots on the $n_{s}$-th sub-slot. $\mathbf{n}_{n_{s}}^{p} = \mathbf{n}_{n_{s}}[:, 1:L_{p}]$.
The CS pilot detection in (9) is a standard multiple-measurement-vectors (MMV) problem, due to the multiple antennas are equipped at the BS.
In this paper, we solve the problem in (9) by the covariance-based ML method \cite{CB-ML}, where the active user number is updated adaptively. 
The $t$-th inner iteration of CB-ML is given by 
\begin{equation} \small
	\begin{aligned}
		& d_{0} \leftarrow {\rm max}\{ \frac{\mathbf{a}^{H}_{t}(\mathbf{\Sigma}^{-1})\hat{\mathbf{\Sigma}}_{y} (\mathbf{\Sigma}^{-1}) \mathbf{a}_{t}- \mathbf{a}^{H}_{t} (\mathbf{\Sigma}^{-1})\mathbf{a}_{t} }{(\mathbf{a}^{H}_{t}\mathbf{\Sigma}^{-1}\mathbf{a}_{t})^{2}}, -\mathbf{\xi}_{t}\} \\
		& \mathbf{\xi}_{t} \leftarrow \mathbf{\xi}_{t}+ d_{0} \\
		& (\mathbf{\Sigma}^{-1}) \leftarrow (\mathbf{\Sigma}^{-1}) - \frac{d_{0} (\mathbf{\Sigma}^{-1}) \mathbf{a}_{t}\mathbf{a}_{t}^{H} (\mathbf{\Sigma}^{-1} ) }{1+d_{0}\mathbf{a}^{H}_{t}(\mathbf{\Sigma}^{-1})\mathbf{a}_{t}   }\\
		& \mathbf{\Sigma} \leftarrow \mathbf{\Sigma}+ d_{0}\mathbf{a}_{t}\mathbf{a}_{t}^{H}, \\
	\end{aligned}
\end{equation}
where $\hat{\mathbf{\Sigma}}_{y} = \frac{1}{M}(\mathbf{Y}_{n_{s}}^{p})^{H}\mathbf{Y}_{n_{s}}^{p}$, $\mathbf{\Sigma} = \mathbf{A}\mathbf{\Gamma}\mathbf{A}^{H}+\sigma^{2}\mathbf{I}_{L_{p}}$, $\mathbf{\Gamma} = {\rm diag}(\mathbf{\xi})$.
The indexes of the CS pilot on the $n_{s}$-th sub-slot is estimated as $\hat{\mathcal{U}}_{n_{s}}^{a} = \{  i \vert \xi_{i} \geq 0, 1\leq i \leq 2^{L_{bp}}  \}$.

Afterward, the CE of the $n_{s}$-th sub-slot can be realized by the minimum mean squared error (MMSE) \cite{fengler2022pilot}, which is given by
\begin{equation}
	\hat{\mathbf{G}}_{n_{s}} = ( \mathbf{A}[:,\hat{\mathcal{U}}_{n_{s}}^{a}])^{T}\mathbf{A}[:,\hat{\mathcal{U}}_{n_{s}}^{a}] )^{-1}(\mathbf{A}[:,\hat{\mathcal{U}}_{n_{s}}^{a}])^{T}\mathbf{Y}_{n_{s}}^{p},
\end{equation}
where $\hat{\mathbf{G}}_{n_{s}}=[\hat{\mathbf{g}}_{1}, \hat{\mathbf{g}}_{2}, \cdots, \hat{\mathbf{g}}_{ \vert  \hat{\mathcal{U}}_{n_{s}}^{a}  \vert }]$.
$N_{a}^{n_{s}} = \vert  \hat{\mathcal{U}}_{n_{s}}^{a}  \vert$ denotes the number of the superposed codewords on the $n_{s}$-th sub-slot.

Based on the estimated $\hat{\mathcal{U}}_{n_{s}}^{a}$, the information bits of the CS part can be decoded by 
\begin{equation}
	\hat{\mathbf{b}}^{p}_{u'} = {\rm bin}( i_{u'}^{p}-1 ), i_{u'}^{p}\in \hat{\mathcal{U}}_{n_{s}}^{a}, 1\leq i_{u'}^{p} \leq 2^{L_{bp}}.
\end{equation}

\subsection{Superposed codeword decomposer}

Inspired by the success in \cite{fengler2022pilot,uncoupledCCS}, 
we can further decompose the superposed codeword of the second data part on the $n_{s}$-th sub-slot by exploiting the spatial diversity provided by the MIMO channel.
With the CE in hand, the superposed signal decomposition is performed in bit-wise.

In more details, the received signal of the $l_{B}$-th ($L_{p}+1 \leq l_{B} \leq L_{bs}+1$) bit on the $n_{s}$-th sub-slot  can be formulated as 
\begin{equation}
	\mathbf{Y}_{n_{s}}[ :, l_{B}  ] = \sum_{   u \in \{ u'\vert \mathbf{s}_{u'}^{AP}[n_{s}]=1\}    } \mathbf{c}_{u}[l_{B}]\mathbf{g}_{u} + \mathbf{n}_{n_{s}}^{d},
\end{equation}
where $\mathbf{n}_{n_{s}}^{d} = \mathbf{n}_{n_{s}}[:, L_{p}+1:L_{bs}+1]$, $\mathbf{c}_{u}[l_{B}]\in\{+1,-1\}$.
$\mathbf{g}_{u}$ is estimated as $\hat{\mathbf{g}}_{u}$ in (11).
In \cite{fengler2022pilot,multiple-stage-orthogonal-pilot}, $\mathbf{c}_{u}[l_{B}]$ is decomposed by the maximum ratio combination (MRC) algorithm with massive MIMO (e.g., $M\geq 30$).
Then, the decomposed single-user bit stream $\hat{\mathbf{c}}_{u}$ is decoded by the single-user polar code decoder.
The inaccuracy of MRC is possibly eliminated by the advanced channel coding.
Unfortunately, MRC suffers a poor performance in our case. There are two main reasons.
First, the spatial diversity is insufficient when the antenna number $M$ is small. 
{For the fair comparison, we assume that the antenna number $M=4$ which is identical to the scheme proposed in \cite{liu2022unsourced}}.
Second, {Similar to \cite{liu2022unsourced}, the channel code is not adopted in the proposed scheme}.
Hence, SCD in our case makes a hard-decision and becomes the bottleneck of our proposal.

To solve this issue, we estimated $\mathbf{c}_{u}[l_{B}], L_{p}<L_{B}<L_{bs}+1$ by the maximum likelihood (ML) detector at the expense of the computational complexity. The decomposition process is formulated as
\begin{equation}\small
	\hat{\mathbf{c}}_{u}[l_{B}] = \mathop{\rm argmin}_{{\mathbf{c}}_{u}[l_{B}] } ||\mathbf{Y}_{n_{s}}[:,l_{B}]-  \sum_{u\in  \{ u'\vert \mathbf{s}_{u'}^{AP}[n_{s}]=1\}   } {\mathbf{c}}_{u}[l_{B}]\hat{\mathbf{g}}_{u}   ||^{2}_{2}.
	\label{ML-SSD}
\end{equation}
Afterwards, $\mathbf{s}_{u}^{\rm BPSK}, \mathbf{s}_{u}^{AP}[n_{s}]=1$ is estimated as $\hat{\mathbf{s}}_{u}^{\rm BPSK}$.
After the BPSK demodulation, $\hat{\mathbf{b}}_{u}^{d}$ is obtained.

\subsection{Pilot-matching-based IM demodulator}

Recall the encoding process in Fig. \ref{system_model}, recovering $\mathbf{b}_{u}^{I}$ is equivalent to detecting  $\mathbf{s}_{u}^{AP}$, which is referred to as IM demodulation.
Since the CS pilot is a part of the codeword, the CS pilot is sent across $K$ different sub-slots.
Thus, we can realize the IM demodulation through the CS pilot matching operation. 
We firstly perform the CS pilot detection over all the sub-slots and obtain $\hat{\mathcal{U}}_{n_{s}}^{a}, 1\leq n_{s} \leq N_{slot}$.
The intersection of these $N_{slot}$ index sets of the CS pilot is the detected channel access pattern sequences.
{For $K=2$,}
the CS pilot matching operation can be formulated as
\begin{equation}
	\hat{\mathbf{s}}_{u}^{AP} \leftarrow \hat{\mathcal{U}}_{n_{s_{1}}}^{a} \cap    \hat{\mathcal{U}}_{n_{s_{2}}}^{a}, 1\leq n_{s1} \neq n_{s2} \leq L_{AP}.
	\label{IM_demodulation}
\end{equation}
With $\hat{\mathbf{b}}_{u}^{p}, \hat{\mathbf{b}}_{u}^{d}, \hat{\mathbf{b}}_{u}^{I}$ in hand, the message $\mathbf{b}_{u}$ can be recovered.

\subsection{SIC and decoding termination criterion}
With $\hat{\mathbf{g}}_{u},\hat{\mathbf{a}}_{i_{u}}, \hat{\mathbf{s}}_{u}^{\rm BPSK}$ and $\hat{\mathbf{s}}_{u}^{AP}$ in hand, the bit stream $\mathbf{b}_{u}$ can be recovered based on formulations (1), (2) and (4).
Then, the corresponding estimated signal is subtracted from the received signal over all the different sub-slots, $\mathbf{Y}$. The SIC process is modeled as 
\begin{equation}
	\mathbf{Y} \leftarrow \mathbf{Y} - \hat{\mathbf{s}}_{u}^{AP} \otimes [\hat{\mathbf{a}}_{i_{u}}, \hat{\mathbf{s}}_{u}^{\rm BPSK}],
\end{equation}
where the subscript `$u$' satisfies the equation (\ref{IM_demodulation}) whose IM demodulation is correctly implemented.

Finally, the decoding termination criterion of the proposed decoder is discussed. The decoder should stop when all the sub-slots become the idle slots.
Hence, the core of decoding termination criterion is the idle slot detection.
To this end, the energy detection in \cite{liu2022unsourced} is invoked.
Let $v = \frac{2}{\sigma^{2}}||\mathbf{Y}_{n_{s}}||_{2}^{2}$.
If the $n_{s}$-th sub-slot is idle, $v$ obeys chi-square distribution with paramater $2(L_{bs}+1)$. Therefore, the detection probability of an idle-sub-slot can be approximated as
\begin{equation}
	\begin{aligned}
		P_{\rm D}^{I} &= P(v\leq \tau_{E}\vert N_{a}^{n_{s}}=0) \\
		& = \Phi( \frac{\tau_{E}-2( L_{bs}+1  ) }{2\sqrt{L_{bs}+1}} ), 
	\end{aligned}
\end{equation}
where $\Phi(\cdot)$ denotes the CDF of a standard Gaussian variable.
As stated in \cite{liu2022unsourced}, if the $n_{s}$-th sub-slot contains more signals, its power will increase and its false alarm probability of idle sub-slot detection decreases ($P(v\leq \tau_{E}\vert N_{a}^{n_{s}}=1 )\geq P(v\leq \tau_{E}\vert N_{a}^{n_{s}}=2)\geq \cdots \geq P(v\leq \tau_{E}\vert N_{a}^{n_{s}}=N_{a}) $).
Accordingly, its upper bound of the false alarm probability is given by
\begin{equation}
	\begin{aligned}
		P_{\rm FA}^{I} &= \sum_{n=1}^{N_{a}} P_{n}P(v \leq \tau_{E} \vert N_{a}^{n_{s}}=n) \\
		& \leq P(v \leq \tau_{E}\vert N_{a}^{n_{s}}=1) \\
		& \approx \Phi(  \frac{  \tau_{E} - 2(L_{bs}+1)(1+{\rm SNR})  }{2( 1+ {\rm SNR}\sqrt{L_{bs}+1} )}  ),
	\end{aligned}
\end{equation}
where $P_{n}$ is the probability that there are $n$ codewords on the $n_{s}$-th sub-slot.
The $\rm SNR$ is computed as $\rm SNR = \mathbb{E}(  \frac{ \sum_{u=1}^{N_{a}} ||\mathbf{g}_{u}^{T}\mathbf{x}_{u}||_{2}^{2}   }{\sigma^{2}(L_{bs}+1)N_{slot}  } )$.
The energy detector design is to choose a proper threshold $\tau_{E}$ which result in $P_{\rm D}^{I}\approx 1$ and $P_{\rm FA}^{I} \approx 0$. The choice of $\tau_{E}$ will be discussed later. 

{Notably, the proposed scheme can work without the knowledge of $N_a$ which is an advantage over the method in \cite{liu2022unsourced}.}

\section{Pilot Collision Resolution and Simplified SCD}
\label{s4}
In this section, the CS pilot collision resolution and the simplified SCD design are further discussed respectively.

\begin{figure}[htbp]
	\centering
	\subfloat[\label{fig-a}]{\includegraphics[scale=0.47]{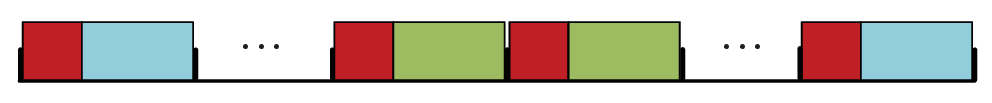}} \\
	\subfloat[\label{fig-b}]{\includegraphics[scale=0.47]{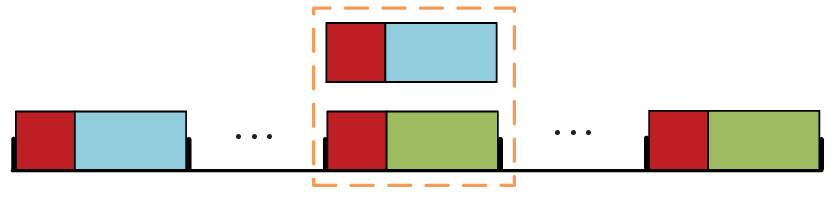}} \\
	\subfloat[\label{fig-c}]{\includegraphics[scale=0.47]{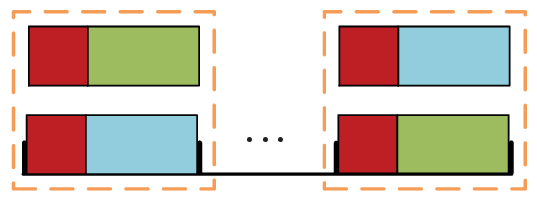}} \\
	\caption{The collision scenario where two different active users select the same CS pilot, where $K=2$: (a) no overlap, (b) partial overlap, (c) complete overlap.}
	\label{pilot_collision}
	\vskip -3mm
\end{figure}

\subsection{CS pilot collision resolution}
Notably, the proposed IM demodulator in (15) can work with the CS pilot collision to some extent. 
Since the CS pilot collision probability is quite low \cite{fengler2022pilot,slotted_pilot}, 
{i.e., the dimension of the CS common codebook $\mathbf{A}$ is extremely large (e.g., $2^{L_{bp}}=2^{14}$) \cite{fengler2022pilot,slotted_pilot}.},
the case where only two active users select the same CS pilot (i.e., $\mathbf{b}_{u_{1}}^{p} = \mathbf{b}_{u_{2}}^{p}, u_{1}\neq u_{2}$) are mainly concerned.
\subsubsection{no overlap:} 
As is shown in Fig. \ref{pilot_collision}-(a), in this case, we can detect a certain CS pilot on four different sub-slots, i.e., $(n_{s_{1}}, n_{s_{2}}, n_{s_{3}}, n_{s_{4}})$-th sub-slots. 
For simplicity, we denote these four codewords as $\hat{\mathbf{c}}_{n_{s_{1}}}, \hat{\mathbf{c}}_{n_{s_{2}}}, \hat{\mathbf{c}}_{n_{s_{3}}}, \hat{\mathbf{c}}_{n_{s_{4}}}$ respectively. 
Also $\hat{\mathbf{c}}_{n_{s_{1}}}^{d}, \hat{\mathbf{c}}_{n_{s_{2}}}^{d}, \hat{\mathbf{c}}_{n_{s_{3}}}^{d}, \hat{\mathbf{c}}_{n_{s_{4}}}^{d}$ denotes the estimation of the BPSK modulated signal part of the codeword.
The similarity of the two codewords can be measured by their Hamming distance.
Let $[i_{1}, i_{2}, i_{3}, i_{4}]$ denote a permutation of $[n_{s_{1}}, n_{s_{2}}, n_{s_{3}}, n_{s_{4}} ]$.
The IM demodulation can be formulated as follows.
\begin{equation}
	[i_{1}, i_{2}, i_{3}, i_{4}] = \mathop{\rm argmin}({\rm hdis}( \hat{\mathbf{c}}_{i_{1}}^{d}, \hat{\mathbf{c}}_{i_{2}}^{d}  ) + {\rm hdis}( \hat{\mathbf{c}}_{i_{3}}^{d}, \hat{\mathbf{c}}_{i_{4}}^{d}  ) ),
\end{equation}
where ${  [i_{1}, i_{2}, i_{3}, i_{4}] \in {\rm Perm}(  n_{s_{1}}, n_{s_{2}}, n_{s_{3}}, n_{s_{4}}   )  }$. $\rm Perm( \mathbf{a} )$ return the full permutation of the vector $\mathbf{a}$ and $\rm hdis(\mathbf{a}, \mathbf{b})$ computes the hamming distance between the vector $\mathbf{a}$ and $\mathbf{b}$.

\subsubsection{partial overlap:}
As is shown in Fig. \ref{pilot_collision}-(b), in this case, we can detect a certain CS pilot on three different sub-slots (e.g., $(n_{s_{1}}, n_{s_{2}}, n_{s_{3}}  )$-th sub-slots). 
Accordingly, there are three codewords detected (i.e., $\hat{\mathbf{c}}_{n_{s_{1}}}, \hat{\mathbf{c}}_{n_{s_{2}}}, \hat{\mathbf{c}}_{n_{s_{3}}}$).
Observe at Fig. \ref{pilot_collision}-(b) that, the core is to identify the sub-slot where two codewords are transmitted. Similar to the no overlap case, we can realize IM demodulation by calculating the similarity of the detected signal on these three sub-slots.
Let $[i_{1}, i_{2}, i_{3}]$ denote a permutation of $[n_{s_{1}}, n_{s_{2}}, n_{s_{3}} ]$.
The IM demodulation can be formulated as follows.
\begin{equation}
	[i_{1}, i_{2}, i_{3}] = \mathop{\rm argmin}||\hat{\mathbf{c}}_{i_{1}}+\hat{\mathbf{c}}_{i_{2}}-\hat{\mathbf{c}}_{i_{3}}||_{2}^{2}.
\end{equation}
\subsubsection{complete overlap:}
Observe Fig. \ref{pilot_collision}-(c) that, in this case, we can detect a certain CS pilot on two different sub-slots.
The channel access pattern sequences of these two codewords are the same. 
The IM demodulation can be performed by (15) successfully.
However, due to the CS pilot collision on the same sub-slot, these two codewords will be termed as a single codeword in the first two decoding phases and a decoding error occurs.

\subsection{Semi-definite relaxation (SDR)-based SCD}

To decompose the superposed codewords on the $n_{s}$-th sub-slot, the ML-based SCD is proposed at the expense of computational complexity in (\ref{ML-SSD}).
Although the ML-based SCD provides a near-optimal performance, it becomes inefficient upon the number of superposed codeword on the same sub-slot increasing.
Hence, designing the simplified SCD is of great significance to cope with the dense active user scenarios.
The problem in (14) can be written as
\begin{equation}
	\hat{\mathbf{c}}_{d} = \mathop{\rm argmin}_{\mathbf{c}_{d}\in\{+1,-1\}^{N_{a}^{n_{s}}  }}||\mathbf{Y}_{n_{s}}[:,l_{B}]-\mathbf{G}_{n_{s}}\mathbf{c}_{d}||_{2}^{2},
	\label{IQP}
\end{equation}
where $\mathbf{c}_{d} = [\mathbf{c}_{1}[l_{B}], \mathbf{c}_{2}[l_{B}], \cdots, \mathbf{c}_{N_{a}^{n_{s}}}[l_{B}] ]$. 
The problem (21) is a typical quadratic programming (QP) with integer constraint, which is generally NP-hard \cite{SDR}. 
Inspired by \cite{SDR}, we utilize the semi-definite relaxation (SDR) technique to solve this problem efficiently.

To apply the SDR, we firstly convert the problem in (\ref{IQP}) into the standard binary quadratic programming (BQP) form. Let 
$\mathbf{g}_{n_{s}} = [ \rm Re\{\mathbf{G}_{n_{s}}\};   \rm Im\{\mathbf{G}_{n_{s}}\}] \in \mathbb{R}^{2M}$, $\mathbf{y}_{n_{s}} =  [ \rm Re\{\mathbf{Y}_{n_{s}}\};   \rm Im\{\mathbf{Y}_{n_{s}}\}]\in \mathbb{R}^{2M}$.
By introducing a slack variable $c\in \{+1,-1\}$,
the optimization problem in (\ref{IQP}) can be rewritten as
\begin{equation}\small
	\begin{aligned}
		& \mathop{\rm min}_{\mathbf{c}_{d}\in \{+1,-1\}^{N_{a}^{n_{s}}}}  \mathbf{c}_{d}\mathbf{g}_{n_{s}}^{T}\mathbf{g}_{n_{s}}\mathbf{c}_{d}-2\mathbf{c}_{d}^{T}\mathbf{g}_{n_{s}}^{T}\mathbf{y}_{n_{s}} \\
		= & \mathop{\rm min}_{\tilde{\mathbf{c}_{d}}\in \{+1,-1\}^{N_{a}^{n_{s}}} }  (c\tilde{\mathbf{c}_{d}})\mathbf{g}_{n_{s}}^{T}\mathbf{g}_{n_{s}}(c\tilde{\mathbf{c}_{d}})-2(c\tilde{\mathbf{c}_{d}})^{T}\mathbf{g}_{n_{s}}^{T}\mathbf{y}_{n_{s}} \\
		= & \mathop{\rm min}_{\tilde{\mathbf{c}_{d}}\in \{+1,-1\}^{N_{a}^{n_{s}}} }  \tilde{\mathbf{c}_{d}}\mathbf{g}_{n_{s}}^{T}\mathbf{g}_{n_{s}}\tilde{\mathbf{c}_{d}}-2(c\tilde{\mathbf{c}_{d}})^{T}\mathbf{g}_{n_{s}}^{T}\mathbf{y}_{n_{s}}\\
		= & \mathop{\rm min}_{\tilde{\mathbf{c}_{d}}\in \{+1,-1\}^{N_{a}^{n_{s}}} } 
		\left[
		\begin{array}{cc}
			\tilde{\mathbf{c}_{d}} \; c
		\end{array}
		\right]
		\left[
		\begin{array}{cc}
			\mathbf{g}_{n_{s}}^{T}\mathbf{g}_{n_{s}} & -\mathbf{g}_{n_{s}}^{T}\mathbf{y}_{n_{s}} \\
			-\mathbf{y}_{n_{s}}^{T}\mathbf{g}_{n_{s}} & 0
		\end{array}
		\right]
		\left[
		\begin{array}{c}
			\tilde{\mathbf{c}_{d}} \\
			c
		\end{array}
		\right] \\
		= & \mathop{\rm min}_{\mathbf{c} \in \{+1,-1\}^{N_{a}^{n_{s}}+1 }  }
		\mathbf{c}^{T}\tilde{\mathbf{G}}\mathbf{c},
	\end{aligned}
\end{equation}
where $ \mathbf{c}^{T} = [\tilde{\mathbf{c}_{d}}, c]  $, $\tilde{\mathbf{G}} = \left[
\begin{array}{cc}
	\mathbf{g}_{n_{s}}^{T}\mathbf{g}_{n_{s}} & -\mathbf{g}_{n_{s}}^{T}\mathbf{y}_{n_{s}} \\
	-\mathbf{y}_{n_{s}}^{T}\mathbf{g}_{n_{s}} & 0
\end{array}
\right] \in \mathbb{R}^{  (N_a^{n_s}+1) \times (N_a^{n_s}+1) }$. 
Since $\mathbf{c}^{T}\tilde{\mathbf{G}}\mathbf{c} = {\rm Tr}(\tilde{\mathbf{G}}\mathbf{c}\mathbf{c}^{T})$, the problem (22) can be simplified as
\begin{equation}
	\begin{aligned}
		\mathop{\rm min} \quad &{\rm Tr}( \tilde{\mathbf{G}}\mathbf{C}  )\\
		s.t. \quad &\mathbf{C}=\mathbf{c}\mathbf{c}^{T} \\ 
		& \mathbf{C}[ n,n ]=1, 1\leq n \leq N_{a}^{n_{s}}
	\end{aligned}
\end{equation}
Note that the constraint `$\mathbf{C}=\mathbf{c}\mathbf{c}^{T}$' is equivalent to the constraint `$\mathbf{C}\succeq 0, \rm rank(\mathbf{C})=1$', where $\rm rank(\cdot)$ returns the rank of a matrix.
Hence, the problem in (23) is rewritten as
\begin{equation}
	\begin{aligned}
		\mathop{\rm min} \quad &{\rm Tr}( \tilde{\mathbf{G}}\mathbf{C}  )\\
		s.t. \quad &\mathbf{C}\succeq 0 \\ 
		& {\rm rank}(\mathbf{C})=1 \\
		& \mathbf{C}[ n,n ]=1, 1\leq n \leq N_{a}^{n_{s}}.
	\end{aligned}
	\label{Tr}
\end{equation}
However, the constraint ${\rm rank}(\mathbf{C})=1$ is non-convex, which makes the problem (24) difficult to solve precisely.
Hence, we may drop this constraint and obtain the SDR of (\ref{Tr}) as follows.
\begin{equation}
	\begin{aligned}
		\mathop{\rm min} \quad &{\rm Tr}( \tilde{\mathbf{G}}\mathbf{C}  )\\
		s.t. \quad &\mathbf{C}\succeq 0 \\ 
		& \mathbf{C}[ n,n ]=1, 1\leq n \leq N_{a}^{n_{s}}.
	\end{aligned}
	\label{SDR}
\end{equation}
Apparently, problem (\ref{SDR}) is convex and can be efficiently solved by the optimization tool CVX \cite{CVX}.
Once the solution of (\ref{SDR}), $\mathbf{C}^{\star}$ is obtained, $150$ candidate solutions of (\ref{Tr}) are generated as recommended by \cite{SDR}.
\begin{itemize}
	\item Generating the $n$-th vector $\hat{\mathbf{c}}^{(n)}, 1\leq n \leq 150$ with length $N_{a}^{n_{s}}$ randomly, which obeys the Gaussian distribution with zero mean and covariance matrix $\mathbf{C}^{\star}$.
	\item Performing the integration over $\hat{\mathbf{c}}^{(n)}$, which is given by
	\begin{equation}
		\hat{\mathbf{c}}^{(n)} = {\rm sgn}( \hat{\mathbf{c}}^{(n)} ),
	\end{equation} 
	where ${\rm sgn}(\cdot)$ is the sign function. 
\end{itemize}
Finally, we get the best approximate solution by 
\begin{equation}
	n^{\star} = \mathop{\rm argmin}_{n=1,\cdots,150} ( \hat{\mathbf{c}}^{(n)} )^{T } \tilde{\mathbf{G}} \hat{\mathbf{c}}^{(n)}. 
\end{equation} 
Based on $\hat{\mathbf{c}}^{(n^{\star})}$, $\mathbf{c}_{d}$ in (\ref{IQP}) can be estimated.

The pseudo-code of the proposed decoder is given in \textbf{Algorithm 1}.

\begin{algorithm}
	\caption{The proposed HD-based decoder}
	\begin{algorithmic}[1]
\REQUIRE{The received signal $\mathbf{Y}$}
\ENSURE{The decoded bit stream $\hat{\mathbf{b}}_{u}, 1\leq u \leq N_{a}$}
\FOR{$1\leq n_{s} \leq N_{slot}$}
	\STATE Compute $\hat{\mathcal{U}}_{n_{s}}^{a}$, $\hat{\mathbf{b}}_{u}^{p}, 1\leq u \leq N_{a}$ and $\hat{\mathbf{G}}_{n_{s}}$ by (11), (12);
\ENDFOR
\WHILE{\rm True} 
	\STATE $n_{s} = \mathop{\rm argmin}_{1 \leq n_{s} \leq N_{slot} }  |\hat{\mathcal{U}}_{n_{s}}^{a}|$; \\
	\IF{$|\hat{\mathcal{U}}_{n_{s}}^{a}| > M$ \rm{or} $|\hat{\mathcal{U}}_{n_{s}}^{a}|==0$ }
		\STATE {Break};
	\ENDIF
		\STATE Compute $\hat{\mathbf{b}}_{u}^{d}, u \in \hat{\mathcal{U}}_{n_{s}}^{a}$ by the  ML-SCD in (14) or the SDR-SCD in (25-27);\\
		\STATE Compute $\hat{\mathbf{b}}_{u}^{I}$ through the CS pilot matching in (15);\\
		\STATE Implement the SIC across the sub-slots by (16);\\
		\STATE Update $\hat{\mathcal{U}}_{n_{s}}^{a}$ and $\hat{\mathbf{G}}_{n_{s}}$ by (10), (11); \\
\ENDWHILE
	\STATE Return $\hat{\mathbf{b}}_{u}, 1\leq u \leq N_{a}$;		
	\end{algorithmic}

\end{algorithm}

\section{Performance Analysis}
\label{s5}
\subsection{The decodable probability analysis}
We first consider the probability that there is at least one decodable sub-slot.
Observe (14) that the superposed codewords on the $n_{s}$-th sub-slot can be properly decomposed in the ideal case when $|\mathcal{U}_{n_{s}}^{a}| \leq M$.
Because $\mathbf{G}_{n_{s}}$ in (21) trends to be full-rank with an overwhelming probability.
Recall that each active user would send its codeword $K$ times based on its access pattern sequence (APs).
At the beginning,
the probability that there are $m, 1\leq m \leq M$ superposed codewords on a single sub-slot is approximately given by
\begin{equation}
	\beta_{m}^{(1)} = \binom{N_{a}}{m}(\frac{K}{N_{slot}})^{m} (1-\frac{K}{N_{slot}})^{N_{a}-m}.
\end{equation}
The probability that there is at least a decodable sub-slot is given by
\begin{equation}
	\gamma_{1} = 1 - ( 1 -  \sum_{m=1}^{M} \beta_{m}^{(1)}  )^{N_{slot}}.
\end{equation}
Assume that only the codewords of a single active user is removed in each decoding iteration, the probability that there are $m, 1\leq m \leq M$ superposed codewords on a single sub-slot within the $t$-th iteration is given by
\begin{equation}
	\beta_{m}^{(t)} = \binom{N_{a}-t}{m}(\frac{K}{N_{slot}})^{m} (1-\frac{K}{N_{slot}})^{N_{a}-t-m}.
\end{equation}
Similarly, the probability that there is at least a decodable sub-slot within the $t$-th iteration is given by
\begin{equation}
	\gamma_{t} = 1 - ( 1 -  \sum_{m=1}^{M} \beta_{m}^{(t)}  )^{N_{slot}}.
\end{equation}
The value of $N_{slot}$ should be selected such that $\gamma_{t}, 1\leq t \leq N_{a}$ trends to be 1 with given $N_{a}$.

\noindent\textbf{Proposition 1:}\emph{When $KN_{a}>MN_{slot}$, $\gamma_{t}$ is a increasing function with respect to $t$ and $N_{slot}$.}

\emph{Proof: See Appendix A.}

Based on the \textbf{Proposition 1}, $N_{slot}$ is selected such that $\gamma_{1}$ trends to be 1.
Recall that the information bits conveyed by the IM approach are $L_{bI}=\lfloor \log_{2}(  \binom{N_{slot}}{K} )    \rfloor$.
Hence, $N_{slot}$ is also selected such that the value of $\vert  \binom{N_{slot}}{K}-2^{L_{bI}}   \vert$ is as small as possible.
To this context, we fix {$K = 2, N_{slot}=33, L_{bI}=9$} in this paper.

\subsection{Throughput analysis}

The throughput $T(r)$ of the proposed scheme can be predicted by
the density evolution method developed in \cite{liu2022unsourced}. $r=\frac{N_{a}}{N_{slot}}$ is the transmission rate. 
The distribution of the transmitted codewords on the sub-slots can be presented by a tanner-graph. 
\begin{figure}[htbp]
	\centering
	\includegraphics[scale=0.81]{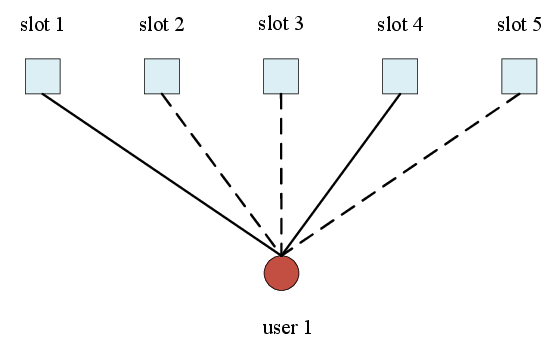}
	\caption{The tanner graph of the spreading step in Fig. \ref{system_model}.}
	\label{tanner-graph}
	\vskip -3mm
\end{figure}
{The tanner graph of the spreading step in Fig. \ref{system_model} is given in Fig. \ref{tanner-graph}}, where the active users are termed as the variable nodes (VNs) and the sub-slots are termed as the check nodes (CNs).
For simplicity, the perfect SIC operation is assumed in our analysis.

In our proposal, each active user would send its codewords on $K$ out of $N_{slot}$ sub-slots.
Hence, the probability that each sub-slot is picked by an active user is approximately $\frac{K}{N_{slot}}$.
Define the right edge distribution $\rho_{j}$ as the proportion of the edges connected to CNs which has degree $j, 1\leq j \le N_{a}$. 
We have $\sum_{j=1}^{N_{a}}\rho_{j} = 1$.
Define the CN distribution $\Pi_{j}$ as the proportion of CNs which has degree $j$.
Then, we have $\rho_{j}KN_{a} = N_{slot}\Pi_{j}j$, which can be further simplified as
\begin{equation}
	\rho_{j} = \frac{\Pi_{j}j}{Kr}.
\end{equation}
Furthermore, $\Pi_{j}$ obeys the binomial distribution ${\rm Binomial}( N_{a}, \frac{K}{N_{slot}} )$.
When $N_{a}\rightarrow \infty, \frac{K}{N_{slot}}  \rightarrow 0$, $\Pi_{j}$ can be approximated as a Poisson distribution ${\rm Poisson}(\frac{KN_{a}}{N_{slot}})$, which is given by
\begin{equation}
	\Pi_{j} = \frac{  (Kr)^{j} \exp(-Kr)  }{  j!  }.
\end{equation}
We have
\begin{equation}
	\rho_{j} = \frac{  (Kr)^{j-1} \exp(-Kr) }{  (j-1)!  }.
\end{equation}
Let $Z_{t}$ denote the probability that an edge is not pruned after the $t$-th iteration.
For simplicity, we first consider the case where the decoder can only decode the sub-slot were $N_{a}^{n_{s}}=1$.
If $N_{a}^{n_{s}}=1$, the $n_{s}$-th sub-slot is referred to as a single-ton.
In this case, the probability that an edge is connected to a CN with degree $1$ in the $t$-th iteration is given by
\begin{equation}
	q_{t}(1) = \sum_{j=1}^{N_{a}} \rho_{j}( 1-Z_{t-1}  )^{j-1}.
\end{equation} 
It implies that $j-1$ edges connected to this CN have been removed in the previous $t-1$ iterations.
A VN would be removed from the graph is at least one of its $K$ edge is connected to a single-ton.
In another words, a edge originated from a VN is not pruned if and only if the other $K-1$ edges of this VN are not pruned.  
With $q_{t}(1)$ in hand, $Z_{t}$ can be computed by
\begin{equation}
	Z_{t}  = (1- q_{t}(1))^{K-1},
\end{equation}
Apparently, (36) is the recursive function of $Z_{t}$, which is initialized as $Z_{0}=1$. Because, no edge would be pruned from the graph at the beginning.

Then, we consider the case where the decoder can decode the sub-slot were $N_{a}^{n_{s}}=2$.
The probability that an edge is connected to a CN with degree $2$ in the $t$-th iteration is given by
\begin{equation}
	q_{t}(2) = \sum_{j=1}^{N_{a}} \rho_{j}  \binom{j-1}{1} Z_{t-1}( 1-Z_{t-1}  )^{j-2}.
\end{equation} 
It means that $j-2$ out of $j-1$ edges connected to this CN is pruned.
Similarly, $Z_{t}$ can be computed as
\begin{equation}
	Z_{t} = (1-q_{t}(1)- q_{t}(2))^{K-1}.
\end{equation}
In our proposal, the number of the superposed codewords on a same sub-slot can be up to $M$ in the idea case.
Based on the above analysis, the probability that an edge is connected to a CN with degree $m$ in the $t$-th iteration is given by
\begin{equation}
	q_{t}(m) = \sum_{j=1}^{N_{a}} \rho_{j}  \binom{j-1}{m-1} Z_{t-1}^{m-1}( 1-Z_{t-1}  )^{j-m}.
\end{equation}
Then, $Z_{t}$ is given by
\begin{equation}
	Z_{t} = (1-\sum_{m=1}^{M}q_{t}(m))^{K-1}.
\end{equation}
Hence, the throughput of the proposed scheme is given by
\begin{equation}
	T(r) = r(1-Z_{t_{max}}),
\end{equation}
where $t_{max}$ denotes the maximal iteration number in the decoding process.
In this paper, the maximal iteration number $t_{max}$ is identical to $N_{a}$.

\subsection{Complexity analysis}
As shown in Fig. \ref{decoder}, the proposed decoder mainly includes CS decoder, SCD, and IM demodulator. 
Since the IM demodulation is realized simply by the CS pilot matching, the overall complexity of the proposed decoder is dominated by the CS decoder and SCD.
As revealed in section V.B, each sub-slot is selected with a probability $\frac{K}{N_{slot}}$ independently.
Hence, the average number of the superposed codeword on a certain sub-slot is $Kr$, where $r=\frac{N_{a}}{N_{slot}}$.
As reported in \cite{CB-ML}, the complexity of CB-ML algorithm is in the order of $\mathcal{O}(  2^{L_{bp}}L_{p}^{2} )$.
The complexity of the CS decoder is $\mathcal{O}(  2^{L_{bp}}L_{p}^{2} N_{slot} )$.
According to \cite{SDR}, the complexity of SDR-based SCD is $\mathcal{O}((Kr)^{3.5}N_{slot}L_{bd})$.
Therefore, the complexity of the proposed decoder is given by
\begin{equation}
	\mathcal{C}_{\rm dec} = 2^{L_{bp}}L_{p}^{2} N_{slot} +  (Kr)^{3.5}N_{slot}L_{bd}.
\end{equation}
Compared with the sparse graph-based URA scheme \cite{liu2022unsourced} whose complexity is in the order of $\mathcal{O}( 2^{Kr}L_{bd} )$, the complexity of the proposed scheme would not increase sharply upon the active user number $N_{a}$ increasing.

\section{Simulation results}
\label{s6}
In this section, the performance of the proposed scheme are evaluated by the exhaustive computer simulations. 
The signal-to-noise ratio (SNR) of each codeword 
{is computed as $\rm SNR = \mathbb{E}(  \frac{ \sum_{u=1}^{N_{a}} ||\mathbf{g}_{u}^{T}\mathbf{x}_{u}||_{2}^{2}   }{\sigma^{2}(L_{bs}+1)N_{slot}  } )$.
	In each coherence interval, $N_a$ bit vector $\{\mathbf{b}_1, \mathbf{b}_2, \cdots, \mathbf{b}_{N_a}\}$ are transmitted.
	For each $\hat{\mathbf{b}}$, if $\hat{\mathbf{b}} \notin \{\mathbf{b}_1, \mathbf{b}_2, \cdots, \mathbf{b}_{N_a}\}$, then, $\hat{\mathbf{b}}$ corresponds to a decoding error, and the frame error rate (FER) is the ratio between the total number of the decoding error and the total number of the transmitted bit vectors.
	The SNR and FER are defined in \cite{liu2022unsourced} and followed in this paper. Furthermore,
}
the normalized squared error (NSE) of the channel estimation is defined as ${\rm NSE} = \frac{  || \hat{\mathbf{G}} - \mathbf{G} ||_{F}^{2} }{ || \mathbf{G} ||_{F}^{2} }$, where $\mathbf{G} = [\mathbf{g}_{1}, \mathbf{g}_{2}, \cdots, \mathbf{g}_{N_{a}}]$.
{The simulation configuration in \cite{liu2022unsourced} is basically followed.
The transmitted bit stream length $L_{bs}= 70$.
The number of antenna number is $M = 4$.
The length of the encoded codeword is $L_{bs}+1 = 71$.
The length of $\mathbf{b}_u^p$, $L_{bp} = 14$ such that the pilot pool is large enough (i.e., $2^{14}$ \cite{fengler2022pilot,slotted_pilot}) to avoiding the pilot collision.
The sub-slot number is $N_{slot} = 33$.
The total channel consumption is $N_{cu} = N_{slot}(L_{bs}+1)$. 
As recommended in \cite{SSC}, $K$ is fixed as 2.
Accordingly, the length of IM bits $L_{bI} = \lfloor  \log_{2}( \binom{N_{slot}}{K} )  \rfloor = 9$.
The length of BPSK data part $L_{bd} = L_{bs} + 1 - L_{bp} - L_{bI} = 48$.
The length of pilot $L_p = L_{bs} + 1 - L_{bd} = 23$. 
}
The system configuration is given in the {Table I}.
\begin{table}[htbp]
	\centering
	\caption{Simulation configuration}
	\begin{tabular}{c|c} 
		\hline 
		Parameters & value \\
		\hline
		The transmitted bit stream length, $L_{bs}$  & 70 \\ 
		\hline
		The CS pilot length $L_p$ &  23 \\
		\hline
		{The length of $\mathbf{b}_{u}^{p}$, $L_{bp}$} & 14 \\
		\hline
		The length of $\mathbf{s}_{u}^{\rm BPSK}$, $L_{bd}$&  48 \\
		\hline
		The sub-slot number, $N_{slot}$ &  33 \\
		\hline
		The value of $K$ & 2\\
		\hline
		The length of IM bits, $L_{bI}$ &  9 \\
		\hline
		The antenna number at the BS, $M$  & 4 \\
		\hline
		The channel use consumption, $N_{cu}$  & 2343 \\
		\hline
	\end{tabular}
\end{table}

\begin{figure}[htbp]
	\centering
	\includegraphics[scale=0.23]{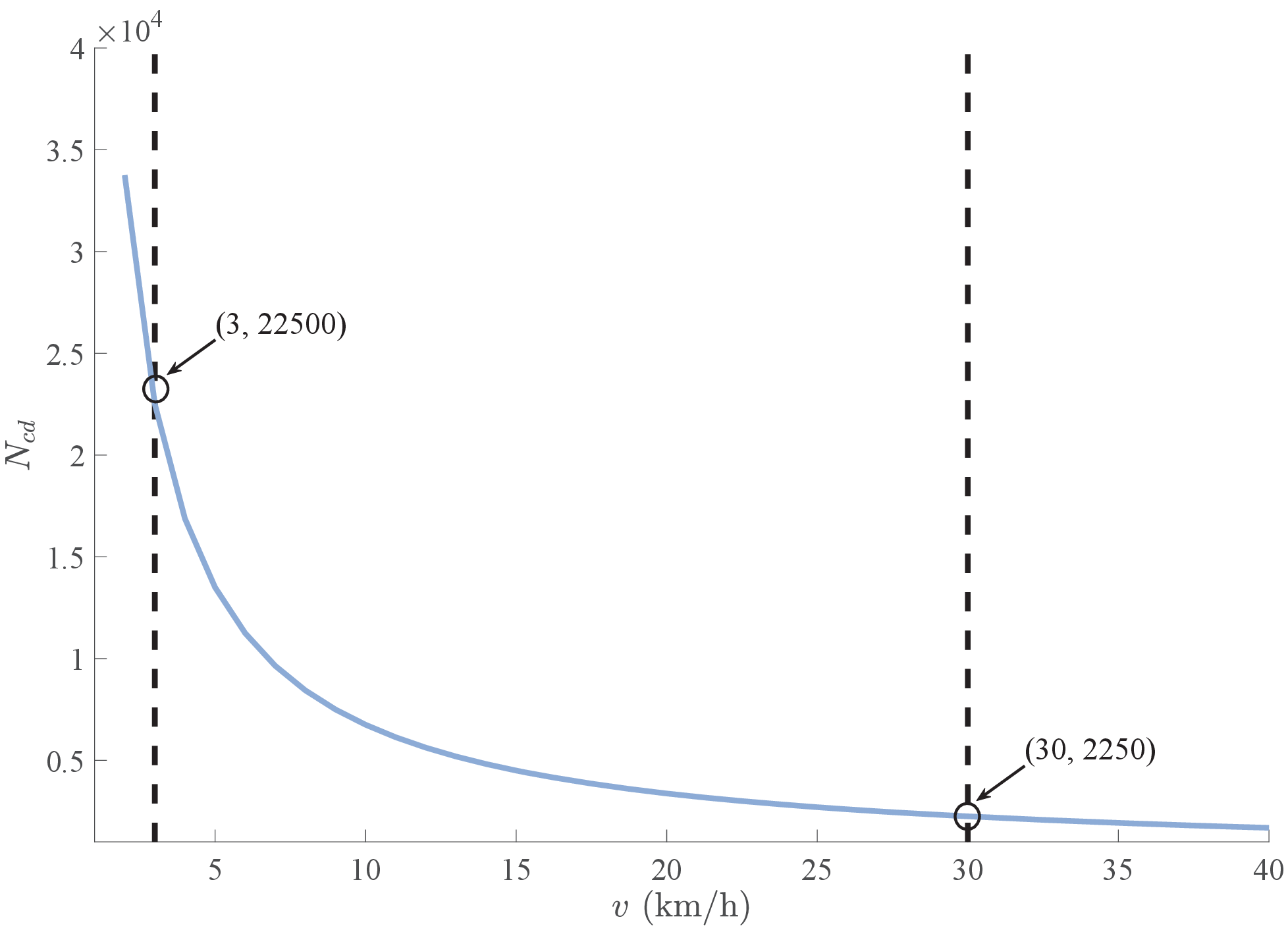}
	\caption{The symbol number of the channel coherent duration, $N_{cd}$ versus velocity $v \rm (km/h)$.}
	\label{channel-coherent-duration}
	\vskip -3mm
\end{figure}

{
	The symbol number of the channel coherent duration, $N_{cd}$ versus different velocity $v \rm (km/h)$ is presented in Fig. \ref{channel-coherent-duration}.
	$N_{cd}$ is computed under the typical wireless configuration, where the carrier frequency is $f_c = 2 \rm GHz$.
	The sampling frequency is chosen in the order of coherence bandwidth, whose typical value is $B_c = 500 \rm KHz$ in outdoor environments.
	The maximal Dopller spread is $f_{max} = \frac{vf_c}{3*10^8}$.
	The coherent time is $T_c = \frac{1}{4f_{max}}$.
	Hence, we have $N_{cd} = B_cT_c$.
	As is shown in Fig. \ref{channel-coherent-duration}, $N_{cd}$ decreases exponentially as the velocity $v$ increases.
	For example, when $v=3 \rm km/h$, $N_{cd}=22500$, which is normally assumed by many existing URA studies e.g., \cite{IntegrateAMP,CCS_list_code,amalladinne2020-original-CCS,T-fold-ALOHA-rayleigh,T-fold-ALOHA-rayleigh,T-fold-IRSA,analysis1-T-fold-IRSA,analysis2-T-fold-IRSA,polar+T-fold,CCS_list_code}. 
	When $v = 30 \rm km/h$, $N_{cd}$ decreases to $2250$. 
	This observation implies that, for the massive Internet-of-thing (IoT) scenarios where devices with a certain velocity (e.g. the agricultural robot, the automated guided vehicle, and the intelligent wearable device), the URA schemes with the large channel coherent duration assumption would become inapplicable.
	Therefore, designing the novel URA scheme under the short channel coherent duration assumption is an open problem, which significantly motivates the reference \cite{liu2022unsourced} and this work.
}

\begin{figure}[htbp]
	\centering
	\includegraphics[scale=0.23]{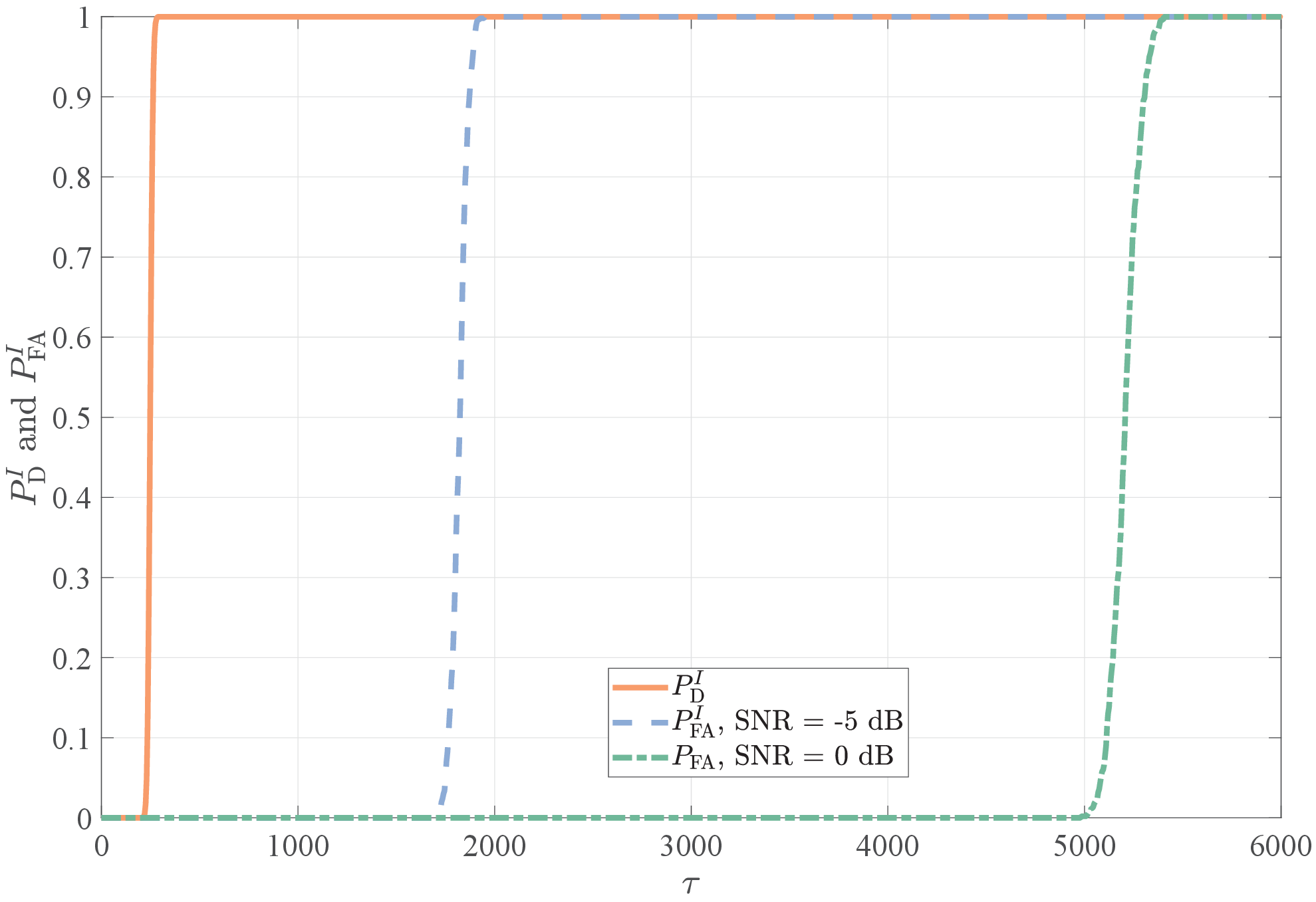}
	\caption{$P_{\rm D}^{I}$ and $P_{\rm FA}^{I}$ versus SNR.}
	\label{energy-detection}
	\vskip -3mm
\end{figure}

The $P_{\rm D}^{I}$ and $P_{\rm FA}^{I}$ versus different $\tau_{E}$ for ${\rm SNR}=-5$ dB and ${\rm SNR}=0$ dB is given in Fig. \ref{energy-detection}.
As is shown in Fig. \ref{energy-detection}, for $\rm SNR=-5$ dB, once the value of $\tau_{E}$ is in the region of $[300, 1500]$, the successful detection probability of the idle sub-slot $P_{\rm D}^{I}$ can get close to 1, meanwhile the false alarm probability $P_{\rm FA}^{I}$ gets close to 0.
Similarly, for $\rm SNR= 0$ dB, the region $[300, 5000]$ is an appropriate region for assigning the value of $\tau_{E}$ to simultaneously satisfy both the detection probability and the false alarm probability requirement.
It reveals that the appropriate regions of $\tau_{E}$ have an intersection for different SNR.
The value within this intersection is a proper threshold.
Based on this principle, in this paper, we set $\tau_{E} = 500$.

\begin{figure}[htbp]
	\centering
	\includegraphics[scale=0.23]{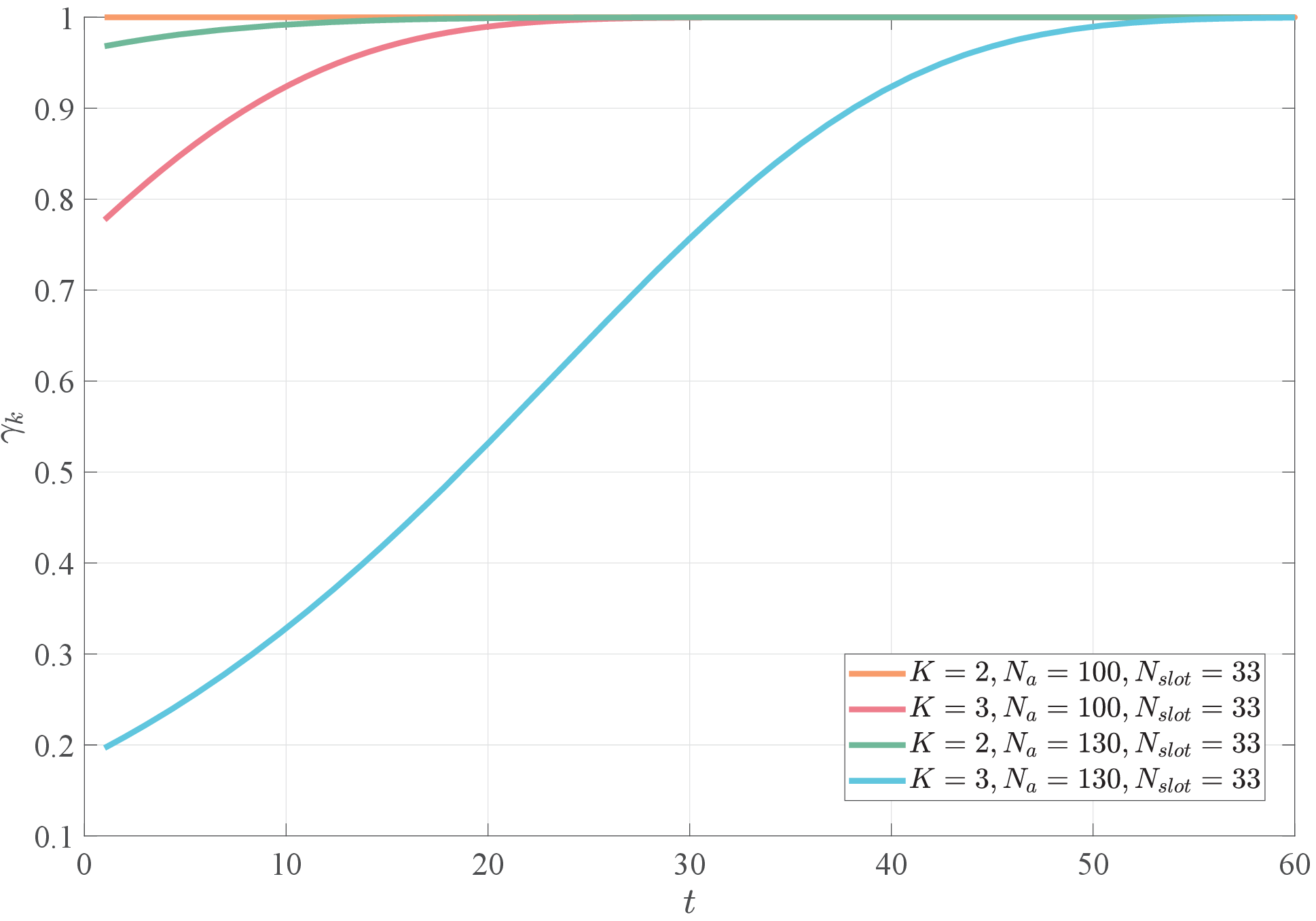}
	\caption{The evolution of the decodable probability $\gamma_{t}$.}
	\label{gammat}
	\vskip -3mm
\end{figure}

The evolution of the decodable probability $\gamma_{t}$ is given in Fig. \ref{gammat}, where the number of the sub-slot is $N_{slot}=33$.
As shown in Fig. \ref{gammat}, $\gamma_{t}$ would increase upon the iteration number $t$ increasing.
This phenomenon verifies the correctness of the \textbf{Proposition 1}.
To implement the slot-wise decoding process successfully, 
we expect that $\gamma_{1}$ approaches 1.
With this principle, the codeword repetition number is more proper to be $K=2$ rather than $K=3$.
Because, $\gamma_{1}$ for $K=2$ is closer to 1 than that for $K=3$, no matter $N_{a}=100$ or $N_{a}= 130$.
This is one of the reasons why we set $K=2$ in this paper.

\begin{figure}[!hb]
	\centering
	\includegraphics[scale=0.23]{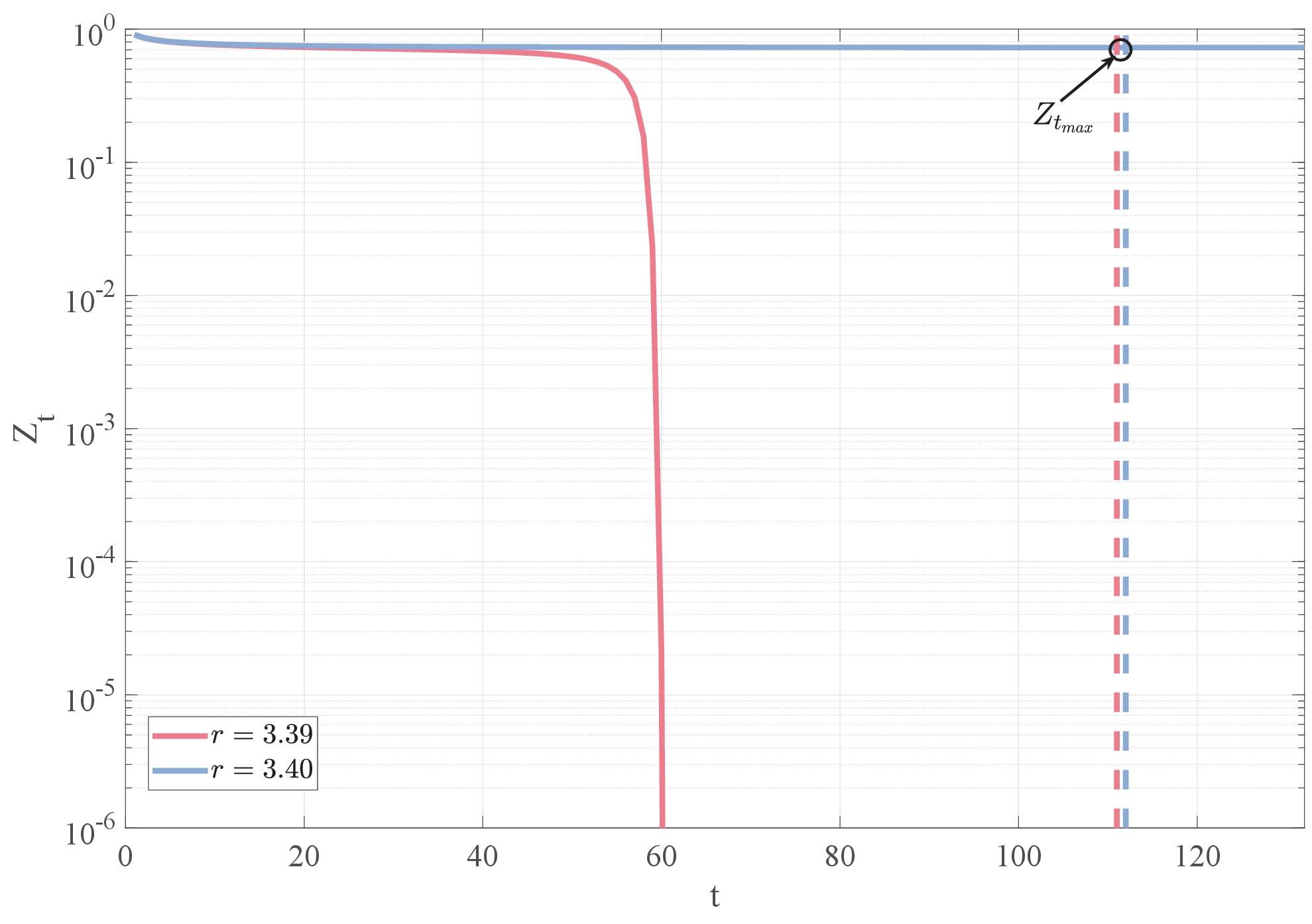}
	\caption{The evolution of the $Z_{t}$.}
	\label{zt}
	\vskip -3mm
\end{figure}

The evolution of the probability that an edge is not pruned from the tanner graph within the $t$-th iteration, $Z_{t}$ is given in Fig. \ref{zt}.
The sub-slot number is $N_{slot}=33$, the codeword repetition number is $K=2$ and the tolerated superposed codeword number on the same sub-slot is $M=4$.
As is shown in Fig. \ref{zt}, there is a threshold of the transmission rate $r$, $r_{th}$. If $r \leq r_{th}$, $Z_{t}$ would converge to 0, otherwise, $Z_{t}$ would not converge to 0.
We find $r_{th}$ by such an iteration approach.
Firstly, we initialize $r = 0.01$.
Then, we compute $Z_{t_{max}}$ based on (38), where the maximal iteration number $t_{max}$ is identical to the active user number $N_{a}$.
If $Z_{t_{max}} <= 10^{-5}$, we would increase $r$ by 0.01 and compute $Z_{t_{max}}$ for one more time. 
Otherwise, the iteration process is terminated. 
In our case, the transmission rate threshold is $r_{th}= 3.39$.

\begin{figure}[!t]
	\centering
	\subfloat[\label{fig-noSIC}]{\includegraphics[scale=0.23]{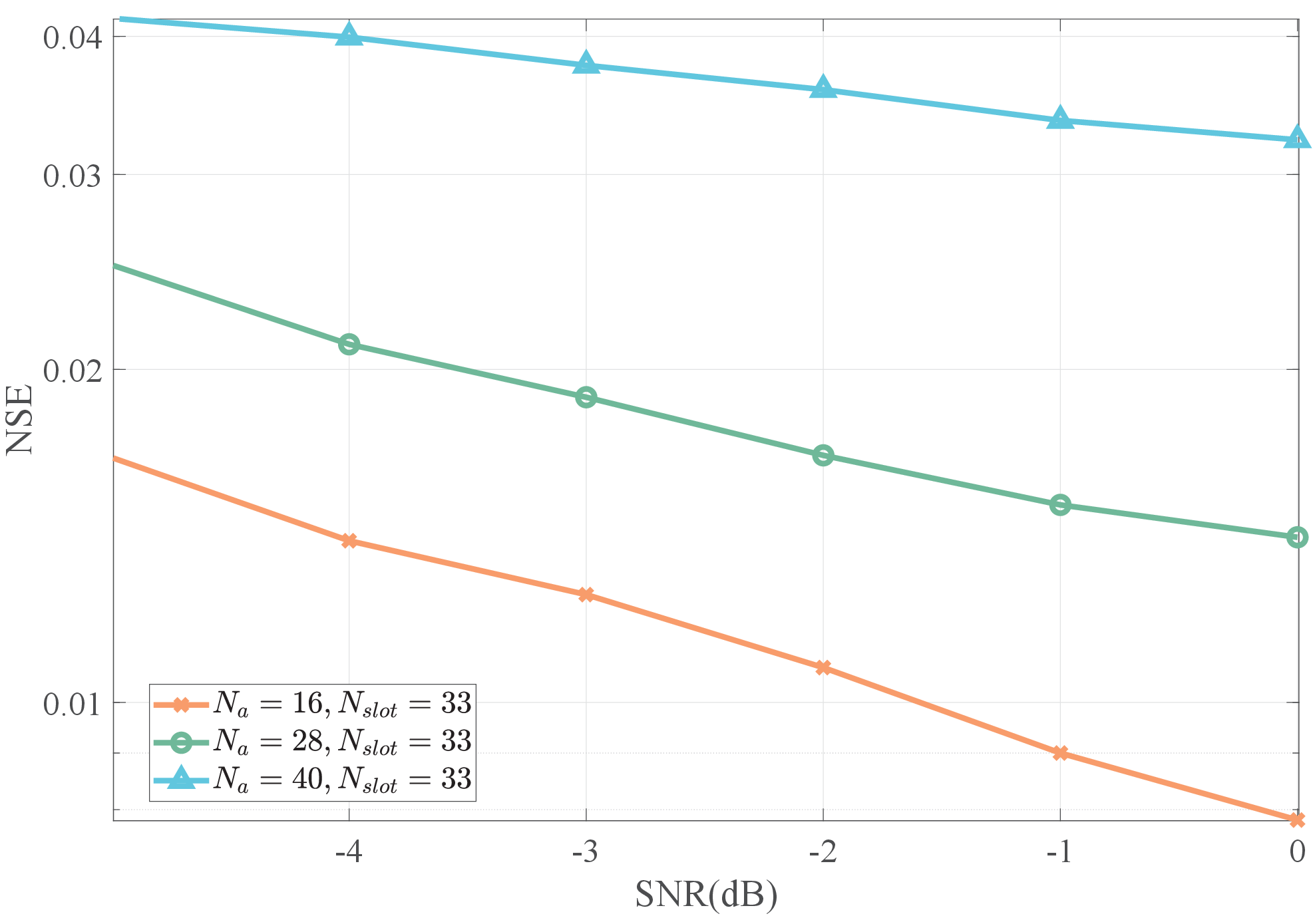}}\\
	\subfloat[\label{fig-SIC}]{\includegraphics[scale=0.23]{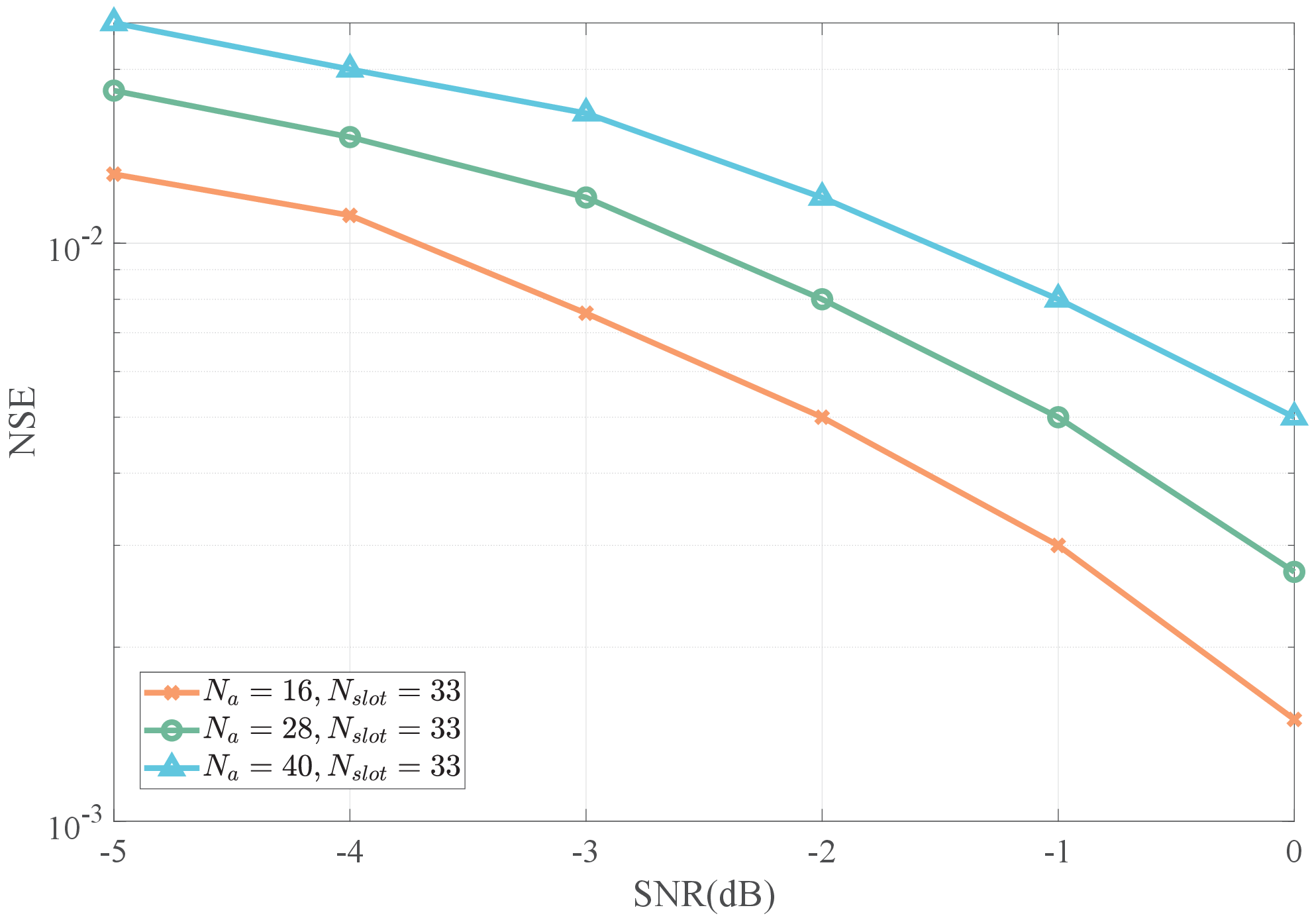}}
	\caption{The NSE performance of the channel estimation: (a) no SIC, (b) SIC.}
	\label{channel_estimation}
\end{figure}
The NSE performance of the CS-based channel estimation is given in \ref{channel_estimation}.
The sub-slot number $N_{slot}=33$, the codeword repetition number $K=2$.
Different active user numbers $N_{a}$ are tested.
As is shown in Fig. \ref{channel_estimation}, the NSE performance would become better upon the SNR increment.
Meanwhile, the increment of the active user number $N_{a}$ would significantly impose the NSE performance, no matter whether SIC operation is considered.
Compare Fig. \ref{channel_estimation}-(a) with Fig. \ref{channel_estimation}-(b), there are two benefits of SIC that can be observed.
First, SIC improves the NSE performance significantly.
Without SIC, the NSE for $N_{a}= 16$ is around $8\times 10^{-3}$ when ${\rm SNR}=0$ dB.
When SIC is considered, the NSE for $N_{a}= 16$ is around $2\times 10^{-3}$ when ${\rm SNR}=0$ dB.
Second, SIC alleviates the impairment brought by the $N_{a}$ increment.
Observe Fig. \ref{channel_estimation}-(a) that the NSE gap between $N_{a}=16$ and $N_{a}=28$ exceeds 2 dB, while this NSE gap is reduced to around 1 dB in Fig. \ref{channel_estimation}-(b).
The results verify the effectiveness of the SIC operation in the proposed decoder.

\begin{figure}[!t]
	\centering
	\includegraphics[scale=0.23]{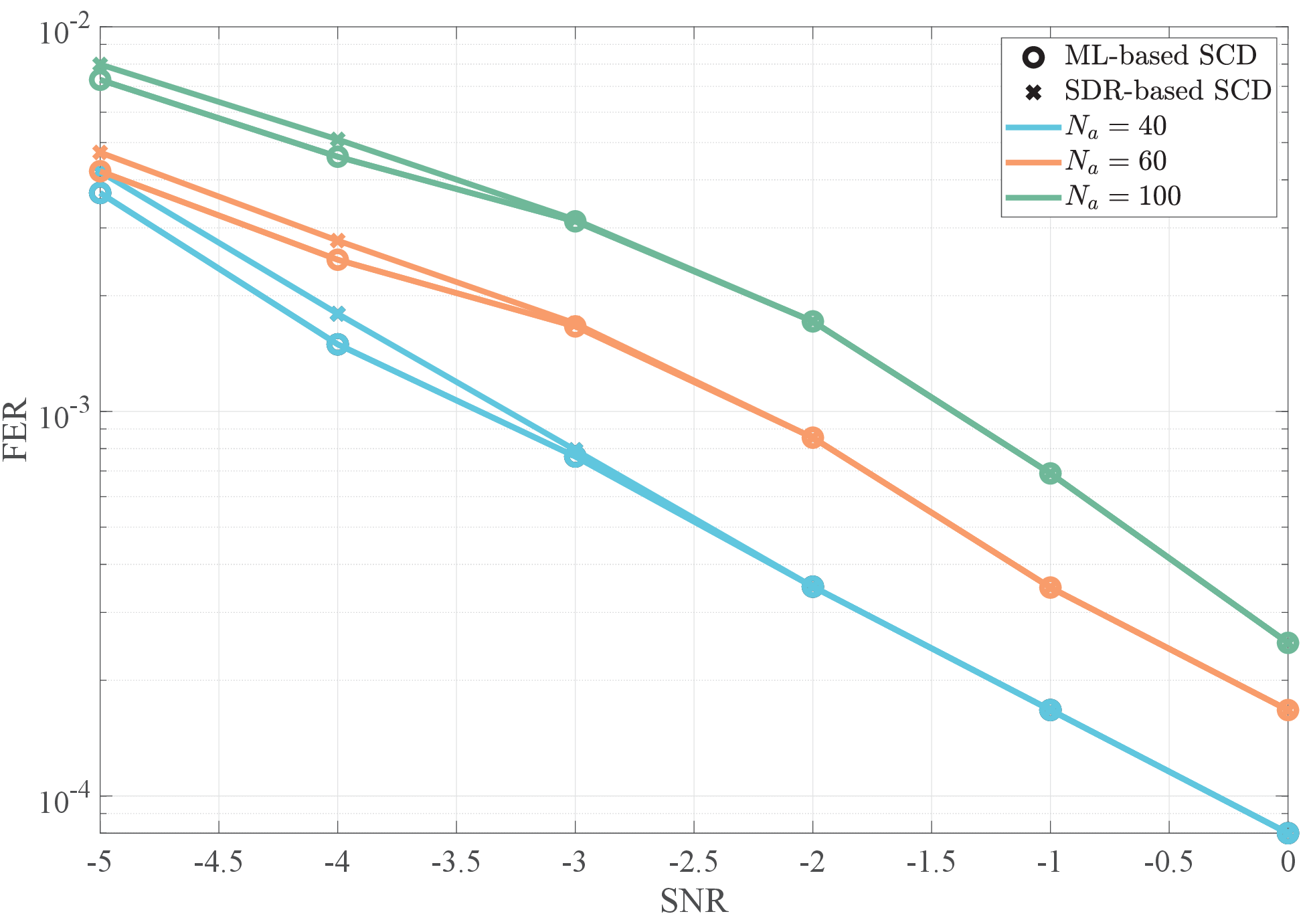}
	\caption{The FER performance versus SNR.}
	\label{FER_SNR}
\end{figure}

The FER performance with different SNRs is simulated in Fig. \ref{FER_SNR}, where different $N_{a}$ are tested. 
The sub-slot number $N_{slot}=33$ and the codeword repetition number $K=2$.
Moreover, the FER performances of the ML-based SCD and the SDR-based SCD are compared.
As is shown in Fig. \ref{FER_SNR}, increasing SNR can significantly improve the FER performance.
Similar to Fig. \ref{channel_estimation}, the increment of $N_{a}$ would impose the FER performance significantly.
Comparing the lines with the circle marker and the lines with the cross marker, only a slight performance degradation can be observed in the low SNR region (e.g. $\rm SNR \leq -3$ dB).
When SNR becomes higher, the FER performances of the ML-based SCD and the SDR-based SCD are almost the same.
This observation confirms the effectiveness of the SDR-based SCD.

\begin{figure}[htbp]
	\centering
	\includegraphics[scale=0.23]{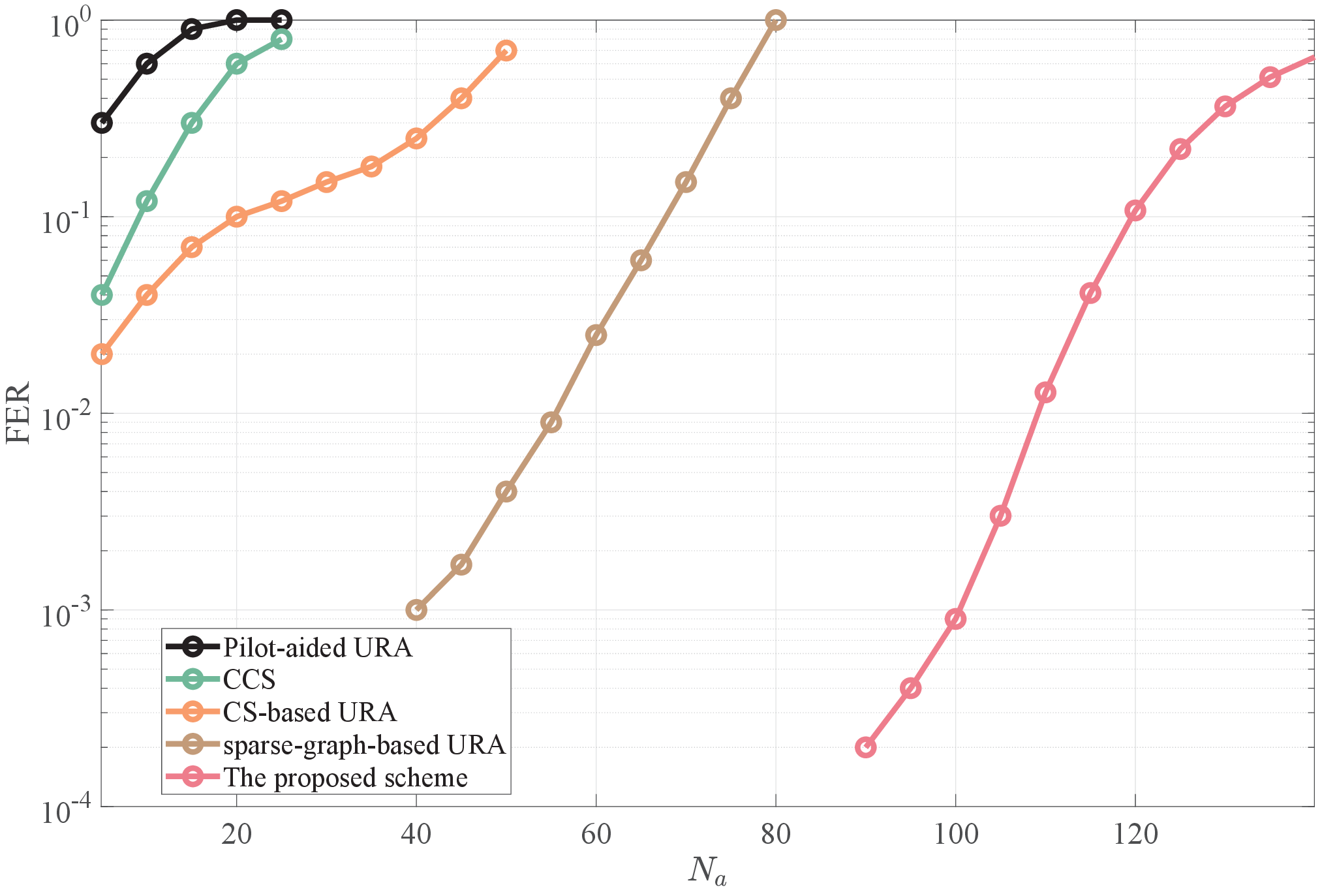}
	\caption{The FER performance versus $N_{a}$.}
	\label{FER_Na}
	\vskip -3mm
\end{figure}

The FER performance versus different $N_{a}$ is presented in Fig. \ref{FER_Na}.
The sub-slot number $N_{slot}=33$ and the codeword repetition number is $K=2$.
The FER performance of CCS is counted,
since the CCS scheme in \cite{amalladinne2020-original-CCS} can work in our slotted framework, where the codeword length is extremely short.
As suggested in \cite{liu2022unsourced}, the information bit steam in CCS is divided into 11 sub-blocks.
The information bit length in the first sub-block is 10 and the information bit length is 6 in the remaining sub-blocks.
The parity-check bit length in the first sub-block is 0 and the parity-check bit length is 4 in the remaining sub-blocks.
Moreover, the CS-based URA in \cite{liu2021sparsity}, where the structure sparsity is exploited to realize the blind decoding, is also considered.
It is noted that although the pilot-aided URA scheme \cite{fengler2022pilot} provides state-of-the-art performance under the massive MIMO channel, this approach suffers a poor FER performance in our system configuration.
The main reason is that the success of the MRC-based SCD in \cite{fengler2022pilot} lies in exploiting the spatial diversity provided by the massive MIMO channel.
When the antenna number $M$ decreases (e.g., $M=4$ in our case), the performance of the MRC-based SCD would decrease rapidly.
Besides, the proposed scheme achieves the best FER performance.
Fixing the target $\rm FER=0.05$, the supported active user number $N_{a}$ of the sparse-graph-based URA \cite{liu2022unsourced} is around 65, while the supported $N_{a}$ of the proposed scheme exceeds 110.
This phenomenon reveals that exploiting the spatial diversity to decompose the superposed codeword on the same sub-slot is more efficient than the peeling decoder.

\begin{figure}[htbp]
	\centering
	\includegraphics[scale=0.23]{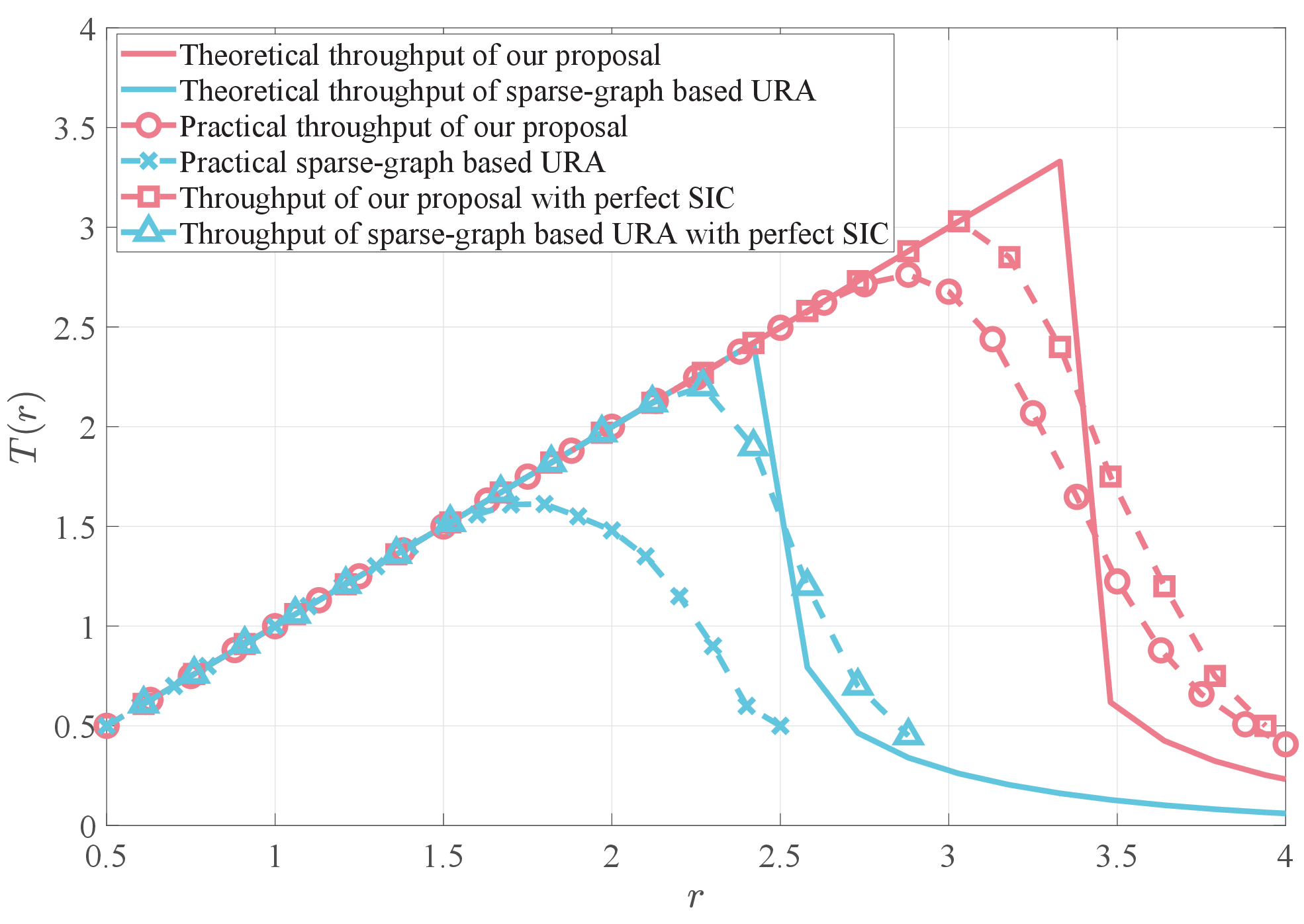}
	\caption{The throughput of the proposed scheme.}
	\label{Tr}
	\vskip -3mm
\end{figure}

The throughput of the proposed scheme is provided in Fig. \ref{Tr}.
{
As is shown in Fig. \ref{Tr}, the theoretical analysis can approximate the throughput of the proposed scheme effectively, especially when $r \leq 3$.
When $r>3$, the analyzed $T(r)$ can still approximately predict the practical throughput.
The result verifies the {effectiveness} of the analysis in (41).
Since more superposed codewords on the same sub-slot can be supported, the throughput of the proposed scheme is much higher than that of the sparse-graph-based URA in \cite{liu2022unsourced}, no matter the theoretical throughput and the practical throughput.
In addition, observing the dashed lines that the imperfect SIC operation would cause significant throughput degradation. Benefiting from a much more accurate channel estimation by compressed sensing (CS) approach, the throughput degradation caused by imperfect SIC of our proposal is much smaller than that of sparse-graph based URA in \cite{liu2022unsourced}.
}

\section{Conclusion}

In this paper, an improved ALOHA-based URA with index modulation is proposed.
The information bits of each active user are divided into three parts which are conveyed by the CS pilot, the BPSK modulation, and the IM respectively.
The CS pilot also serves the consequent channel estimation.
Accordingly, a hard-decision-based decoder including the CS decoder, ML-based SCD, and the IM demodulator is further developed to implement slot-wise data decoding.
To further reduce the complexity, an SDR-based SCD is considered.
Moreover, the throughput analysis of the proposed scheme is conducted by the density evolution tool. 
The exhaustive computer simulation results verify the superiority of the proposed scheme.
{Since the short channel coherent duration is considered, the proposed scheme is more applicable to the IoT scenarios where the devices with some velocity ($v\leq 30 \rm km/h$).}

{
Our future work includes improving the system performance of our proposal by adopting the advanced index modulation strategies \cite{IM1,IM2}.
}

\appendix
Firstly, we would like to show that $\gamma_{t}$ is an increasing function with respect to $t$.
To this end, we can compare $\gamma_{1}$ with $\gamma_{2}$ simply, which is equivalent to compare $\beta_{m}^{(1)}, 1\leq m \leq M$ with $\beta_{m}^{(2)}, 1\leq m \leq M$.
Let $r_{0} = \frac{ \beta_{m}^{(1)}  }{  \beta_{m}^{(2)}  }$, we have
\begin{equation}
	\begin{aligned}
		r_{0} &= \frac{   \binom{N_{a}}{m}(\frac{K}{N_{slot}})^{m} (1-\frac{K}{N_{slot}})^{N_{a}-m}     }{   \binom{N_{a}-1}{m}(\frac{K}{N_{slot}})^{m} (1-\frac{K}{N_{slot}})^{N_{a}-m-1}    } \\
		&= (\frac{N_{a}}{N_{a}-m})(1- \frac{K}{N_{slot}})\\
		&= \frac{    1- \frac{K}{N_{slot}}      }{   1- \frac{m}{N_{a}}.             }
	\end{aligned}
\end{equation}
With $KN_{a} > MN_{slot} $ in hand, we have $r_{0} <1$.
It implies $\beta_{m}^{(2)} > \beta_{m}^{(1)}$ and $\gamma_{2}> \gamma_{1}$ is yielded. 

Then, we would show that $\gamma_{t}$ is a increasing function with respect to $N_{slot}$.
The partial derivative of $\beta_{m}^{(t)}$ with respect to $N_{slot}$ is given by
\begin{equation} \footnotesize
	\begin{aligned}
		\frac{  \partial \beta_{m}^{(t)}   }{     \partial N_{slot}        } 
		&= \binom{N_{a}-t}{m} \frac{K}{N_{slot}^{2}}c^{m-1} (1-c)^{N_{a}-t-m-1}
		(-m+\frac{N_{a}K}{N_{slot}}) \\
		& \propto KN_{a}-mN_{slot},
	\end{aligned}
\end{equation}
where $c=\frac{K}{N_{slot}}$.
With $KN_{a}> mN_{slot}$ in hand, we have $\frac{  \partial \beta_{m}^{(t)}   }{     \partial N_{slot}        } > 0$.
Since $1-\sum_{m=1}^{M}\beta_{m}^{(t)}<1$, $\gamma_{t}$ would increase upon $N_{slot}$ increasing.
The proof is accomplished.

\bibliographystyle{gbt7714-numerical}
\bibliography{myref}

\begin{thebibliography}{39}
\providecommand{\natexlab}[1]{#1}
\providecommand{\url}[1]{#1}
\expandafter\ifx\csname urlstyle\endcsname\relax\else
  \urlstyle{same}\fi
\expandafter\ifx\csname href\endcsname\relax
  \DeclareUrlCommand\doi{\urlstyle{rm}}
  \def\eprint#1#2{#2}
\else
  \def\doi#1{\href{https://doi.org/#1}{\nolinkurl{#1}}}
  \let\eprint\href
\fi

\bibitem[Bockelmann et~al.(2018)Bockelmann, Pratas, Wunder, Saur, Navarro,
  Gregoratti, Vivier, De~Carvalho, Ji, Stefanovi{\'c},
  et~al.]{bockelmann2018towards}
BOCKELMANN C, PRATAS N~K, WUNDER G, et~al.
\newblock Towards massive connectivity support for scalable
  \textcolor{blue}{mMTC} communications in {5G} networks\allowbreak[J].
\newblock IEEE \textcolor{blue}{Access}, 2018, 6: 28969-28992.

\bibitem[Chen et~al.(2018)Chen, Sohrabi, and Yu]{chen2018sparse}
CHEN Z, SOHRABI F, YU W.
\newblock Sparse activity detection for massive connectivity\allowbreak[J].
\newblock IEEE Transactions on Signal Processing, 2018, 66\allowbreak (7):
  1890-1904.

\bibitem[Liu et~al.(2018)Liu and Yu]{liu2018massive}
LIU L, YU W.
\newblock Massive connectivity with massive {MIMO}-{Part I}: Device activity
  detection and channel estimation\allowbreak[J].
\newblock IEEE Transactions on Signal Processing, 2018, 66\allowbreak (11):
  2933-2946.

\bibitem[Liu et~al.(2018)Liu, Larsson, Yu, Popovski, Stefanovic, and
  De~Carvalho]{liu2018sparse}
LIU L, LARSSON E~G, YU W, et~al.
\newblock Sparse signal processing for grant-free massive connectivity: A
  future paradigm for random access protocols in the internet of
  things\allowbreak[J].
\newblock IEEE Signal Processing Magazine, 2018, 35\allowbreak (5): 88-99.

\bibitem[Polyanskiy(2017)]{Polyanskiy-perspective}
POLYANSKIY Y.
\newblock {A} perspective on massive random-access\allowbreak[C]//\allowbreak
2017 IEEE International Symposium on Information Theory (ISIT).
\newblock 2017: 2523-2527.

\bibitem[Ordentlich et~al.(2017)Ordentlich and
  Polyanskiy]{original-T-fold-ALOHA}
ORDENTLICH O, POLYANSKIY Y.
\newblock Low complexity schemes for the random access {Gaussian}
  channel\allowbreak[C]//\allowbreak
2017 IEEE International Symposium on Information Theory (ISIT).
\newblock 2017: 2528-2532.

\bibitem[Kowshik et~al.(2020)Kowshik, Andreev, Frolov, and
  Polyanskiy]{T-fold-ALOHA-rayleigh}
KOWSHIK S~S, ANDREEV K, FROLOV A, et~al.
\newblock Energy efficient coded random access for the wireless
  uplink\allowbreak[J].
\newblock IEEE Transactions on Communications, 2020, 68\allowbreak (8):
  4694-4708.

\bibitem[Ustinova et~al.(2019)Ustinova, Glebov, Rybin, and Frolov]{T-fold-IRSA}
USTINOVA D, GLEBOV A, RYBIN P, et~al.
\newblock Efficient concatenated same codebook construction for the random
  access {Gaussian MAC}\allowbreak[C]//\allowbreak
2019 IEEE 90th Vehicular Technology Conference (VTC2019-Fall).
\newblock 2019: 1-5.

\bibitem[Glebov et~al.(2019)Glebov, Matveev, Andreev, Frolov, and
  Turlikov]{analysis1-T-fold-IRSA}
GLEBOV A, MATVEEV N, ANDREEV K, et~al.
\newblock Achievability bounds for {T-Fold} irregular repetition slotted
  {ALOHA} scheme in the {Gaussian MAC}\allowbreak[C]//\allowbreak
2019 IEEE Wireless Communications and Networking Conference (WCNC).
\newblock 2019: 1-6.

\bibitem[Ustinova et~al.(2019)Ustinova, Rybin, and
  Frolov]{analysis2-T-fold-IRSA}
USTINOVA D, RYBIN P, FROLOV A.
\newblock On the analysis of {T-Fold} coded slotted {ALOHA} for a fixed error
  probability\allowbreak[C]//\allowbreak
2019 11th International Congress on Ultra Modern Telecommunications and Control
  Systems and Workshops (ICUMT).
\newblock 2019: 1-5.

\bibitem[Vem et~al.(2019)Vem, Narayanan, Chamberland, and
  Cheng]{CS+sparse-spreading}
VEM A, NARAYANAN K~R, CHAMBERLAND J~F, et~al.
\newblock A user-independent successive interference cancellation based coding
  scheme for the unsourced random access {Gaussian} channel\allowbreak[J].
\newblock IEEE Transactions on Communications, 2019, 67\allowbreak (12):
  8258-8272.

\bibitem[Andreev et~al.(2020)Andreev, Marshakov, and Frolov]{polar+T-fold}
ANDREEV K, MARSHAKOV E, FROLOV A.
\newblock A polar code based {TIN-SIC} scheme for the unsourced random access
  in the quasi-static fading {MAC}\allowbreak[C]//\allowbreak
2020 IEEE International Symposium on Information Theory (ISIT).
\newblock 2020: 3019-3024.

\bibitem[Amalladinne et~al.(2020)Amalladinne, Chamberland, and
  Narayanan]{amalladinne2020-original-CCS}
AMALLADINNE V~K, CHAMBERLAND J~F, NARAYANAN K~R.
\newblock A coded compressed sensing scheme for unsourced multiple
  access\allowbreak[J].
\newblock IEEE Transactions on Information Theory, 2020, 66\allowbreak (10):
  6509-6533.

\bibitem[Fengler et~al.(2021)Fengler, Jung, and Caire]{fengler2021sparcs}
FENGLER A, JUNG P, CAIRE G.
\newblock {SPARCs} for unsourced random access\allowbreak[J].
\newblock IEEE Transactions on Information Theory, 2021, 67\allowbreak (10):
  6894-6915.

\bibitem[Amalladinne et~al.(2021)Amalladinne, Pradhan, Rush, Chamberland, and
  Narayanan]{IntegrateAMP}
AMALLADINNE V~K, PRADHAN A~K, RUSH C, et~al.
\newblock Unsourced random access with coded compressed sensing: Integrating
  {AMP} and belief propagation\allowbreak[J].
\newblock IEEE Transactions on Information Theory, 2021, 68\allowbreak (4):
  2384-2409.

\bibitem[Cao et~al.(2023)Cao, Xing, and Liang]{cao2023crc}
CAO H, XING J, LIANG S.
\newblock {CRC}-aided sparse regression codes for unsourced random
  access\allowbreak[J].
\newblock IEEE Communications Letters, 2023.

\bibitem[Andreev et~al.(2022)Andreev, Rybin, and Frolov]{CCS_list_code}
ANDREEV K, RYBIN P, FROLOV A.
\newblock Coded compressed sensing with list recoverable codes for the
  unsourced random access\allowbreak[J].
\newblock IEEE Transactions on Communications, 2022, 70\allowbreak (12):
  7886-7898.

\bibitem[Jiang et~al.(2023)Jiang and Fan]{jiang2023raptor}
JIANG D, FAN P.
\newblock A raptor code based unsourced random access with coordinated
  tree-raptor decoding algorithm\allowbreak[C]//\allowbreak
2023 IEEE/CIC International Conference on Communications in China (ICCC
  Workshops).
\newblock IEEE, 2023: 1-6.

\bibitem[Shyianov et~al.(2020)Shyianov, Bellili, Mezghani, and
  Hossain]{uncoupledCCS}
SHYIANOV V, BELLILI F, MEZGHANI A, et~al.
\newblock Massive unsourced random access based on uncoupled compressive
  sensing: Another blessing of massive {MIMO}\allowbreak[J].
\newblock IEEE Journal on Selected Areas in Communications, 2020, 39\allowbreak
  (3): 820-834.

\bibitem[Xie et~al.(2022)Xie, Wu, An, Gao, Zhang, Xing, Wong, and
  Xiao]{CCSangulardomain}
XIE X, WU Y, AN J, et~al.
\newblock Massive unsourced random access: Exploiting angular domain
  sparsity\allowbreak[J].
\newblock IEEE Transactions on Communications, 2022, 70\allowbreak (4):
  2480-2498.

\bibitem[Che et~al.(2022)Che, Zhang, Yang, Chen, Zhong, and Ng]{CCSbeamspace}
CHE J, ZHANG Z, YANG Z, et~al.
\newblock Unsourced random massive access with beam-space tree
  decoding\allowbreak[J].
\newblock IEEE Journal on Selected Areas in Communications, 2022, 40\allowbreak
  (4): 1146-1161.

\bibitem[Wang et~al.(2021)Wang, Zhang, Chen, Zhong, and Hanzo]{CCSRM}
WANG J, ZHANG Z, CHEN X, et~al.
\newblock Unsourced massive random access scheme exploiting reed-muller
  sequences\allowbreak[J].
\newblock IEEE Transactions on Communications, 2021, 70\allowbreak (2):
  1290-1303.

\bibitem[Fengler et~al.(2022)Fengler, Musa, Jung, and Caire]{fengler2022pilot}
FENGLER A, MUSA O, JUNG P, et~al.
\newblock Pilot-based unsourced random access with a massive {MIMO} receiver,
  interference cancellation, and power control\allowbreak[J].
\newblock IEEE Journal on Selected Areas in Communications, 2022, 40\allowbreak
  (5): 1522-1534.

\bibitem[Ahmadi et~al.(2023)Ahmadi, Kazemi, and
  Duman]{multiple-stage-orthogonal-pilot}
AHMADI M~J, KAZEMI M, DUMAN T~M.
\newblock Unsourced random access using multiple stages of orthogonal pilots:
  {MIMO} and single-antenna structures\allowbreak[J].
\newblock IEEE Transactions on Wireless Communications, 2023.

\bibitem[Gkagkos et~al.(2023)Gkagkos, Narayanan, Chamberland, and
  Georghiades]{FASURA}
GKAGKOS M, NARAYANAN K~R, CHAMBERLAND J~F, et~al.
\newblock \textcolor{blue}{FASURA}: A scheme for quasi-static fading unsourced
  random access channels\allowbreak[J].
\newblock IEEE Transactions on Communications, 2023.

\bibitem[Ozates et~al.(2023)Ozates, Kazemi, and Duman]{slotted_pilot}
OZATES M, KAZEMI M, DUMAN T~M.
\newblock A slotted pilot-based unsourced random access scheme with a
  multiple-antenna receiver\allowbreak[J].
\newblock IEEE Transactions on Wireless Communications, 2023: 1-1.

\bibitem[Liu et~al.(2022)Liu and Wang]{liu2022unsourced}
LIU J, WANG X.
\newblock Unsourced multiple access based on sparse tanner graph—efficient
  decoding, analysis, and optimization\allowbreak[J].
\newblock IEEE Journal on Selected Areas in Communications, 2022, 40\allowbreak
  (5): 1509-1521.

\bibitem[Ustinova et~al.(2022)Ustinova, Frolov, and Andreev]{polar+T-fold_MIMO}
USTINOVA D, FROLOV A, ANDREEV K.
\newblock Unsourced random access pilot-assisted polar code construction for
  {MIMO} channel\allowbreak[C]//\allowbreak
2022 IEEE International Multi-Conference on Engineering, Computer and
  Information Sciences (SIBIRCON).
\newblock 2022: 1-4.

\bibitem[Mao et~al.(2021)Mao and Wang]{mao2021terahertz}
MAO T, WANG Z.
\newblock Terahertz wireless communications with flexible index modulation
  aided pilot design\allowbreak[J].
\newblock IEEE Journal on Selected Areas in Communications, 2021, 39\allowbreak
  (6): 1651-1662.

\bibitem[Mao et~al.(2024)Mao, Zhou, Xiao, Han, and Wang]{mao2024index}
MAO T, ZHOU Z, XIAO Z, et~al.
\newblock Index-modulation-aided terahertz communications with reconfigurable
  intelligent surface\allowbreak[J].
\newblock IEEE Transactions on Wireless Communications, 2024.

\bibitem[Xiao et~al.(2023)Xiao, Liu, Mao, Liu, Zhang, Xia, and
  Hanzo]{xiao2023twin}
XIAO Z, LIU G, MAO T, et~al.
\newblock Twin-layer ris-aided differential index modulation dispensing with
  channel estimation\allowbreak[J].
\newblock IEEE Transactions on Vehicular Technology, 2023.

\bibitem[Yang et~al.(2023)Yang and Fan]{yang2023improved}
YANG L, FAN P.
\newblock Improved sparse vector code based on optimized spreading matrix for
  short-packet in urllc\allowbreak[J].
\newblock IEEE Wireless Communications Letters, 2023.

\bibitem[Zhang et~al.(2022)Zhang, Zhang, Shim, Han, Zhang, and Sato]{SSC}
ZHANG X, ZHANG D, SHIM B, et~al.
\newblock Sparse superimposed coding for short-packet urllc\allowbreak[J].
\newblock IEEE Internet of Things Journal, 2022, 9\allowbreak (7): 5275-5289.

\bibitem[Fengler et~al.(2021)Fengler, Haghighatshoar, Jung, and Caire]{CB-ML}
FENGLER A, HAGHIGHATSHOAR S, JUNG P, et~al.
\newblock Non-bayesian activity detection, large-scale fading coefficient
  estimation, and unsourced random access with a massive {MIMO}
  receiver\allowbreak[J].
\newblock IEEE Transactions on Information Theory, 2021, 67\allowbreak (5):
  2925-2951.

\bibitem[Luo et~al.(2010)Luo, Ma, So, Ye, and Zhang]{SDR}
LUO Z~Q, MA W~K, SO A~M~C, et~al.
\newblock Semidefinite relaxation of quadratic optimization
  problems\allowbreak[J].
\newblock IEEE Signal Processing Magazine, 2010, 27\allowbreak (3): 20-34.

\bibitem[Guimaraes et~al.(2015)Guimaraes, Floriano, and Chaves]{CVX}
GUIMARAES D~A, FLORIANO G~H~F, CHAVES L~S.
\newblock A tutorial on the \textcolor{blue}{CVX} system for modeling and
  solving convex optimization problems\allowbreak[J].
\newblock IEEE Latin America Transactions, 2015, 13\allowbreak (5): 1228-1257.

\bibitem[Liu et~al.(2021)Liu and Wang]{liu2021sparsity}
LIU J, WANG X.
\newblock Sparsity-exploiting blind receiver algorithms for unsourced multiple
  access in {MIMO} and massive {MIMO} channels\allowbreak[J].
\newblock IEEE Transactions on Communications, 2021, 69\allowbreak (12):
  8055-8067.

\bibitem[Xiao et~al.(2020)Xiao, Chen, Hemadeh, Xiao, and Jiang]{IM1}
XIAO L, CHEN D, HEMADEH I~A, et~al.
\newblock Graph theory assisted bit-to-index-combination gray coding for
  generalized index modulation\allowbreak[J].
\newblock IEEE Transactions on Wireless Communications, 2020, 19\allowbreak
  (12): 8232-8245.

\bibitem[Xiao et~al.(2023)Xiao, Zhai, Liu, Liu, Xiao, and Jiang]{IM2}
XIAO L, ZHAI X, LIU Y, et~al.
\newblock A unified bit-to-symbol mapping for generalized constellation
  modulation\allowbreak[J].
\newblock China Communications, 2023, 20\allowbreak (6): 229-239.

\end{thebibliography}


\end{document}